%% file: arxiv_movm.tex
\documentclass[10pt,draftclsnofoot,onecolumn]{IEEEtran}
\usepackage{amsmath}
\usepackage{amssymb}
\usepackage{upgreek}
\usepackage{tipa}
\usepackage{enumerate}
\usepackage{subfig}
\usepackage{float}
\usepackage{multicol}
\usepackage[usenames,dvipsnames]{pstricks}
\usepackage{epsfig}
\usepackage{pst-grad} % For gradients
\usepackage{pst-plot} % For axes
\usepackage[noadjust]{cite}
\usepackage{caption}
\usepackage{graphicx}
\usepackage{epstopdf}
\usepackage{psfrag}
\usepackage{amsfonts}
\usepackage{color}
\usepackage{amsmath,mathtools}
\usepackage{psfrag}
\usepackage{comment}
\usepackage{url}
\usepackage{balance}

\begin{document}
\title{The Modified Optimal Velocity Model: \\ Stability Analyses and Design Guidelines}
%\title{A computational comparison of the optimal velocity model and the modified optimal velocity model} 

\author{\IEEEauthorblockN{Gopal Krishna Kamath, Krishna Jagannathan and Gaurav Raina} \\
\IEEEauthorblockA{Department of Electrical Engineering, Indian Institute of Technology Madras, Chennai 600 036, India\\
Email: $\lbrace \text{ee12d033, krishnaj, gaurav} \rbrace$@ee.iitm.ac.in}
\thanks{A conference version of this paper appeared in Proceedings of the $53^{rd}$ Annual Allerton Conference on Communication, Control and Computing, pp. 538-545, 2015. DOI: 10.1109/ALLERTON.2015.7447051}
}
 %\author{Gopal~Krishna~Kamath, Krishna~Jagannathan and Gaurav~Raina % <-this % stops a space
%\thanks{G.K.~Kamath, K.~Jagannathan and G.~Raina are with the Department
%of Electrical Engineering, Indian Institute of Technology Madras, Chennai 600 036, India.  
%E-mail: \{ee12d033, krishnaj, gaurav\}@ee.iitm.ac.in}
%\thanks{A conference version of this paper appeared in Proceedings of the $53^{rd}$ Annual Allerton Conference on Communication, Control and Computing, pp. 538-545, 2015. DOI: 10.1109/ALLERTON.2015.7447051}}
%\markboth{IEEE Transactions on Automatic Control}%
%{G.K.~Kamath \MakeLowercase{\textit{et al.}}: The reduced classical car-following model: stability analyses and design guidelines }
\maketitle

\begin{abstract}
Reaction delays are important in determining the qualitative dynamical properties of a platoon of vehicles traveling on a straight road. In this paper, we investigate the impact of delayed feedback on the dynamics of the Modified Optimal Velocity Model (MOVM). Specifically, we analyze the MOVM in three regimes -- no delay, small delay and arbitrary delay. In the absence of reaction delays, we show that the MOVM is locally stable. For small delays, we then derive a sufficient condition for the MOVM to be locally stable. Next, for an arbitrary delay, we derive the necessary and sufficient condition for the local stability of the MOVM. We show that the traffic flow transits from the locally stable to the locally unstable regime via a Hopf bifurcation. We also derive the necessary and sufficient condition for non-oscillatory convergence and characterize the rate of convergence of the MOVM. These conditions help ensure smooth traffic flow, good ride quality and quick equilibration to the uniform flow. Further, since a Hopf bifurcation results in the emergence of limit cycles, we provide an analytical framework to characterize the type of the Hopf bifurcation and the asymptotic orbital stability of the resulting non-linear oscillations. Finally, we corroborate our analyses using stability charts, bifurcation diagrams, numerical computations and simulations conducted using MATLAB.
\end{abstract}


\IEEEpeerreviewmaketitle

\section{Introduction}
\label{sec:intro}
\input{sec1_intro.tex}

\section{Models}
\label{sec:models}
\input{sec2_models.tex}

\section{The no-delay regime}
\label{sec:ideal}
\input{sec3_ideal.tex}

\section{The small-delay regime}
\label{sec:small}
\input{sec4_small.tex}

\section{The arbitrary-delay regime}
\label{sec:hopf}
\input{sec5_hopf.tex}

\section{Non-oscillatory convergence}
\label{sec:noc}
\input{sec6_noc.tex}

\section{Rate of convergence}
\label{sec:roc}
\input{sec7_roc.tex}

\section{Hopf bifurcation analysis}
\label{section:RCCFMhopf}
\input{sec8_hba.tex}

\section{Simulations}
\label{sec:sims}
\input{sec9_sims.tex}

\section{Concluding Remarks}
\label{sec:conc}
\input{sec10_conc.tex}

\section*{Acknowledgements}
This work is undertaken as a part of an Information Technology Research Academy (ITRA), Media Lab Asia, project titled ``De-congesting India's transportation networks.'' The authors are also thankful to Debayani Ghosh, Rakshith Jagannath and Sreelakshmi Manjunath for many helpful discussions.

\end{document}

%% file: sec1_intro.tex
Intelligent transportation systems constitute a substantial theme of discussion on futuristic smart cities. In this context, self-driven vehicles are a prospective solution to address traffic issues such as resource utilization and commute delays; see~\cite[Section 5.2]{RR},~\cite{AV, SG, VAC} and references therein. To ensure that these objectives are met, in addition to ensuring human safety, the design of control algorithms for these vehicles becomes important. To that end, it is imperative to have an in-depth understanding of human behavior and vehicular dynamics. This has led to the development and study of a class of dynamical models known as the car-following models~\cite{DCG, MBD, DC, DH, GO, GKK}.

Feedback delays play an important role in determining the qualitative behavior of dynamical systems~\cite{HL}. In particular, these delays are known to destabilize the system and induce oscillatory behavior~\cite{GKK,RS}. In the context of human-driven vehicles, predominant components of the reaction delay are psychological and mechanical in nature~\cite{RS}. In contrast, the delay in self-driven vehicles arise due to sensing, communication, signal processing and actuation, and are envisioned to be smaller than human reaction delays~\cite{AK}.

%An important consideration in the study of car-following models is the delay in the dynamical variables. Delays arise due to various factors such as sensing, mechanical motions, communication  and signal processing. These delays are known to have a variety of effects on the properties of a dynamical system~\cite{HL}. Specifically, delays can readily lead to oscillations and instability~\cite{RS, XZ}.

In this paper, we investigate the impact of delayed feedback on the qualitative dynamical properties of a platoon of vehicles traveling on a straight road. Specifically, we consider each vehicle's dynamics to be modeled by the Modified Optimal Velocity Model (MOVM)~\cite{GKK}. Motivated by the wide range of values assumed by reaction delays in various scenarios, we analyze the MOVM in three regimes; namely, $(i)$ no delay, $(ii)$ small delay and $(iii)$ arbitrary delay. In the absence of delays, we show that the MOVM is locally stable. When the delays are rather small, as in the case of self-driven vehicles, we derive a sufficient condition for the local stability of the MOVM using a suitable approximation. For the arbitrary-delay regime, we analytically characterize the region of local stability for the MOVM.

In the context transportation networks, two additional properties are of practical importance; namely, ride quality (lack of jerky vehicular motion) and the time taken by the platoon to attain the desired equilibrium when perturbed. Mathematically, these translate to studying the non-oscillatory property of the MOVM's solutions and the rate of their convergence to the desired equilibrium. In this paper, we also characterize these properties for the MOVM.

In the context of human-driven vehicles, model parameters generally correspond to human behavior, and hence cannot be ``tuned'' or ``controlled.'' However, our work enhances phenomenological insight into the emergence and evolution of traffic congestion. For example, a peculiar phenomenon known as the ``phantom jam'' is observed on highways~\cite{DC,DH}. Therein, a congestion wave emerges seemingly out of nowhere and propagates up the highway from the point of its origin. Such an oscillatory behavior in the traffic flow has typically been attributed to a change in the driver's sensitivity, such as a sudden deceleration; for details, see~\cite{DC,DH}. In general, feedback delays are known to induce oscillations in state variables of dynamical systems~\cite{GKK,RS}. Since the MOVM explicitly incorporates feedback delays, and relative velocities and headways constitute state variables of the MOVM, our work provides a theoretical basis for understanding the emergence and evolution of oscillatory phenomena such as ``phantom jams.'' In particular, our work serves to highlight the possible role of reaction delays in the emergence of oscillatory phenomena in traffic flows. More generally, our results reveal an important observation: the traffic flow may transit into instability due to an appropriate variation in \emph{any} subset of model parameters. To capture this complex dependence of stability on various parameters, we introduce an exogenous, non-dimensional parameter in our dynamical model. We then analyze the behavior of the resulting system as the exogenous parameter is pushed just beyond the stability boundary. We show that non-linear oscillations, termed \emph{limit cycles}, emerge in the traffic flow due to a \emph{Hopf bifurcation}.

%In the context of human-driven vehicles, our investigation into the impact of reaction delay enhances phenomenological insight into the emergence and evolution of traffic congestion. For example, a peculiar phenomenon known as a ``phantom jam'' -- the emergence of a back-propagating congestion wave in motorway traffic, seemingly out of nowhere -- has been observed in the real world~\cite{DC,DH}. Previous studies~\cite{DC,DH} have shown that a change in driver's sensitivity (for instance, a sudden deceleration) can lead to such oscillatory behavior. In this paper, we show that similar oscillations could also result from an increase in the driver's reaction delay. More generally, our study leads to an important observation that the transition of traffic flow from stability to instability could take place due to a variation in many combinations of model parameters. In order to capture this complex dependence on various parameters, we introduce an exogenous, non-dimensional parameter in our dynamical model. We then analyze the system behavior as this exogenous parameter pushes the system just beyond the stability boundary, and show that limit cycles emerge due to a certain bifurcation exhibited by the dynamical system.

In the context of self-driven vehicles, reaction delays are expected to be smaller than their human counterparts~\cite{AK}. Hence, it would be realistically possible to achieve smaller equilibrium headways~\cite[Section 5.2]{RR}. This would, in turn, vastly improve resource utilization without compromising safety~\cite{SG}. In this paper, based on our theoretical analyses, we provide some design guidelines to appropriately tune the parameters of the so-called ``upper longitudinal control algorithm''~\cite[Section 5.2]{RR}. Mathematically, our analytical findings highlight the quantitative impact of delayed feedback on the design of control algorithms for self-driven vehicles. Specifically, our design guidelines take into consideration various aspects of the longitudinal control algorithm such as stability, good ride quality and fast convergence of the traffic to the uniform flow. In the event that the traffic flow does lose stability, our design guidelines help tune the model parameters with an aim of reducing the amplitude and angular velocity of the resultant limit cycles.

%The impact of the reaction delay is perhaps even more important in the context of self-driven vehicles. Self-driven vehicles are envisioned to have reduced reaction delays as compared to a human driver. As a result, self-driven vehicles facilitate smaller equilibrium separation between consecutive vehicles~\cite[Section 5.2]{RR}. This, in turn, improves resource utilization without compromising safety~\cite{SG}.  In contrast to the case of human-driven vehicles, the parameters in the control algorithm -- known as upper longitudinal control algorithm~\cite[Section 5.2]{RR} -- for self-driven vehicles need to be tuned appropriately. To that end, our analytical findings highlight the quantitative impact of delayed feedback on the design of control algorithms for self-driven vehicles. In particular, the combination of stability, convergence and detailed bifurcation analyses may help in the design of various aspects of longitudinal control algorithms. We complement our theoretical analyses using bifurcation diagrams and MATLAB simulations.

\subsection{Related work on car-following models}   

The motivation for our paper comes from the key idea behind the Optimal Velocity Model (OVM) proposed by Bando \emph{et al.} in~\cite{MB}. However, the model considered therein was devoid of reaction delays. Thus, a new model was proposed in~\cite{MBD} to account for the drivers' delays. Therein, the authors also claimed that these delays were not central to capturing the dynamics of the system. In response, Davis showed via numerical computations that reaction delays indeed played an important part in determining the qualitative behavior of the OVM~\cite{DCMBD}. This led to a further modification to the OVM in~\cite{LCD}. However, this too did not account for the delay arising due to a vehicle's own velocity.

It was shown in~\cite{IG} that the OVM without delays loses local stability via a Hopf bifurcation. For the OVM with delays,~\cite{GO1} performed an initial numerical study of the bifurcation phenomenon before supplying an analytical proof in~\cite{GO}. The theme issue on ``Traffic jams: dynamics and control''~\cite{GO2} highlights the growing interest in a  dynamical systems viewpoint of transportation networks. A recent exposition of linear stability analysis in the context of car-following models can be found in~\cite{REW}.

From a vehicular dynamics perspective, most upper longitudinal controllers in the literature assume the lower controller's dynamics to be well modeled by a first-order control system, in order to capture the delay lag~\cite[Section 5.3]{RR}. The upper longitudinal controllers are then designed to maintain either constant velocity, spacing or time gap; for details, see~\cite{RR1} and the references therein. Specifically, it was shown in~\cite{RR1} that synchronization with the lead vehicle is possible by using information only from the vehicle directly ahead. This reduces implementation complexity, and does not mandate vehicles to be installed with communication devices.

However, in the context of autonomous vehicles, communication systems are required to exchange various system states required for the control algorithm. This information is used either for distributed control~\cite{RR1} or coordinated control~\cite{ZQ}. Formation and platoon stabilities have also been studied considering information flow among the vehicles~\cite{RUC,THS}. For an extensive review, see~\cite{KCD}. Usage of communication systems is also known to mitigate phantom jams~\cite{MTS}.

At a microscopic level, Chen \emph{et al.} proposed a behavioral car-following model based on empirical data that captures phantom jams~\cite{DC1}. Therein, the authors showed statistical correlation in drivers' behavior before and during traffic oscillations. However, no suggestions to avoid phantom jams were offered. To that end, Nishi \emph{et al.} developed a framework for ``jam-absorbing'' driving in~\cite{RN}. A ``jam-absorbing vehicle'' appropriately varies its headway with the aim of mitigating phantom jams. This work was extended by Taniguchi \emph{et al.}~\cite{YT} to include car-following behavior. Therein, the authors also numerically constructed the region in parameter space that avoids formation of secondary jams.

%this paper considers the problem of stabilizing a platoon of autonomous vehicles. In contrast, 

In the context of platoon stability, it has been shown that well-placed, communicating autonomous vehicles may be used to stabilize platoons of human-driven vehicles~\cite{GO3}. More generally, the platooning problem has been studied as a consensus problem with delays~\cite{MAS}. Such an approach aids the design of coupling protocols between interacting agents (in this context, vehicles). In contrast, we provide design guidelines to appropriately choose protocol parameters, for a given coupling protocol.

%Gong \emph{et al.} achieve platoon stability by developing a controller using an optimization approach~\cite{SG1}. A key idea is to code the stability conditions into the constraints of the optimization problem. This approach requires the availability of global information, and hence suffers from two specific drawbacks: $(i)$ it may be hard to obtain such information, and $(ii)$ the overhead to obtain it may be significant. In comparison, our approach requires information only from the vehicle directly ahead. It is relatively easy to obtain while incurring minimal possible delay.

\subsection{Our contributions}
Our contributions are as follows.
\begin{itemize}
\item[(1)] We derive a variant of the OVM for an infinitely-long road -- called the MOVM -- and analyze it in three regimes; namely, $(i)$ no delay, $(ii)$ small delay and $(iii)$ arbitrary delay. We prove that the ideal case of instantaneously-reacting drivers is locally stable for all practically significant parameter values. We then derive a stability condition for the small-delay regime by conducting a linearization on the time variable.
\item[(2)] For the case of an arbitrary delay, we derive the necessary and sufficient condition for the local stability of the MOVM. We then prove that, upon violation of this condition, the MOVM loses local stability via a Hopf bifurcation.
\item[(3)] We provide an analytical framework to characterize the type of the Hopf bifurcation and the asymptotic orbital stability of the emergent limit cycles using Poincar\'{e} normal forms and the center manifold theory.
\item[(4)] In the case of human-driven vehicles, our work enhances phenomenological insight into the emergence and evolution of traffic congestion. For example, the Hopf bifurcation analysis provides a mathematical framework to offer a possible explanation for the observed ``phantom jams''~\cite{GKK}. In the case of self-driven vehicles, our work offers suggestions for their design guidelines.
\item[(5)] We derive a necessary and sufficient condition for non-oscillatory convergence of the MOVM. This is useful in the context of a transportation network since oscillations lead to jerky vehicular movements, thereby degrading ride quality and possibly causing collisions.
\item[(6)] We characterize the rate of convergence of the MOVM, thereby gaining insight into the time required for the platoon to equilibrate, when perturbed. Such perturbations occur, for instance, when a vehicle departs from a platoon. Therein, we also bring forth the trade-off between the rate of convergence and non-oscillatory convergence of the MOVM.
%\item[(6)] We highlight the trade-off between non-oscillatory convergence and the rate of convergence. Considering this trade-off, we suggest some guidelines to appropriately choose parameters for the upper longitudinal control algorithm in self-driven vehicles.
\item[(7)] We corroborate the analytical results with the aid of stability charts, bifurcation diagrams, numerical computations and simulations performed using MATLAB.
\end{itemize}

%A part of this work appeared in the conference version~\cite{GKK} of this paper, while point $(5)$ is taken from~\cite{GKKITS}. Remaining points listed above enhances our understanding of the MOVM.

%In summary, in this work, we investigate the impact of delayed feedback on the qualitative dynamical properties of a platoon of vehicles driving on a straight road. Specifically, we focus on analysing the effect of delayed feedback on the RCCFM in three regimes; namely, $(i)$ no delay, $(ii)$ small delay and $(iii)$ arbitrary delay. Furthermore, we find the condition that ensures smooth traffic flow, and also characterise the rate at which the RCCFM equilibrates. Finally, we corroborate our analytical findings using stability charts, bifurcation diagrams and MATLAB simulations.

The remainder of this paper is organized as follows. In Section~\ref{sec:models}, we summarize the OVM and derive the MOVM. In Sections~\ref{sec:ideal},~\ref{sec:small} and~\ref{sec:hopf}, we characterize the stable regions for the MOVM in no-delay, small-delay and arbitrary-delay regimes respectively. We then derive the necessary and sufficient condition for non-oscillatory convergence of the MOVM in Section~\ref{sec:noc}, and characterize its rate of convergence in Section~\ref{sec:roc}. In Section~\ref{section:RCCFMhopf}, we present the local Hopf bifurcation analysis for the MOVM. In Section~\ref{sec:sims}, we corroborate our analyses using MATLAB simulations before concluding the paper in Section~\ref{sec:conc}.

%% file: sec2_models.tex
In this section, we first provide an overview of the setting of our work. We then briefly explain the OVM, before ending the section by deriving the MOVM.

\subsection{The setting}
We consider $N+1$ idealistic vehicles (with $0$ length) traveling on an infinitely long, single-lane road with no overtaking. The lead vehicle is indexed with $0,$ the vehicle following it with $1,$ and so on. The acceleration of each vehicle is updated based on a combination of its position, velocity and acceleration as well as those corresponding to the vehicle directly ahead. We use $x_i(t),$ $\dot{x}_i(t)$ and $\ddot{x}_i(t)$ to denote the position, velocity and acceleration of the vehicle indexed $i$  at time $t$ respectively. We also assume that the lead vehicle's acceleration and velocity profiles are known. Specifically, we only consider leader profiles that converge to $\ddot{x}_0 = 0$ and $0 <$ $\dot{x}_0 < \infty$ in finite time; that is, there exists $T_0 < \infty$ such that $\ddot{x}_0(t) = 0,$ $\dot{x}_0(t) = \dot{x}_0 > 0,$ $\forall \, t \geq T_0$. We also use the terms ``driver'' and ``vehicle'' interchangeably throughout.

\subsection{The Optimal Velocity Model (OVM)}
\label{subsec:OVM}

The OVM, proposed by Bando \emph{et al.} in~\cite{MB}, is based on the key idea that each vehicle in a platoon tries to attain an ``optimal'' velocity, which a function of its headway. Hence, each vehicle updates its acceleration proportional to the difference between this optimal velocity and its own velocity. This was modified in~\cite{MBD} to account for the reaction delay. For $N$ vehicles traveling on a circular loop of length $L$ units, the dynamics is captured by~\cite{MBD}
\begin{align}
\nonumber
\ddot{x}_1(t) = & \, a \left( V( x_{N}(t - \tau) - x_1(t - \tau)) - \dot{x}_1(t - \tau) \right), \\ \label{eq:OVMO}
\ddot{x}_i(t) = & \, a \left( V( x_{i-1}(t - \tau) - x_i(t - \tau)) - \dot{x}_i(t - \tau) \right),
\end{align}
for $i \in \lbrace 2, \cdots, N \rbrace.$ Here, $a > 0$ is the drivers' sensitivity coefficient, $\tau$ is the common reaction delay and $V: \mathbb{R}_+ \rightarrow \mathbb{R}_+$ is called the Optimal Velocity Function (OVF). As pointed out in~\cite{MBET}, an OVF satisfies:
\begin{itemize}
\item[(i)] Monotonic increase,
%, $i.e.$, $ y_2 > y_1 \implies V(y_2) > V(y_1) $.
\item[(ii)] Bounded above, and,
%, $i.e.$, $\exists$ $M$ such that $0 < M < \infty$ and $V(y) < M \text{ } \forall y$.
\item[(iii)] Continuous differentiability.
\end{itemize}
Let $V^{\max} = \underset{y \rightarrow \infty}{\text{lim}} V(y).$ The limit exists as a consequence of (i) and (ii) above. Also, (iii) ensures that an OVF will also be invertible.

\subsection{The Modified Optimal Velocity Model (MOVM)}

%A reason we wish to better understand the dynamics on straight roads is the numerical study~\cite{GKKITS}, where it was shown that periodic boundary conditions may induce instabilities in the OVM.

Next, we derive a variant of the OVM for our setting. To that end, we define the relative spacing (or headway) and the relative velocity as $y_i(t)$ = $x_{i-1}(t) - x_{i}(t)$ and $v_i(t)$ = $\dot{y}_i(t)$ = $\dot{x}_{i-1}(t) - \dot{x}_{i}(t)$ respectively. Substituting these in~\eqref{eq:OVMO}, we obtain the following system after some algebraic manipulations
\begin{align}
\nonumber
\dot{v}_1(t) = & \, \ddot{x}_0(t) + a \left( \dot{x}_0(t - \tau_1) -  V(y_1(t - \tau_1)) - v_1(t - \tau_1) \right), \\ \nonumber
\dot{v}_k(t) = & \, a \left( V(y_{k-1}(t - \tau_{k-1})) - V(y_{k}(t - \tau_k)) - v_k(t - \tau_k) \right), \\ \label{eq:MOVM123}
\dot{y}_i(t) = & \, v_i(t),
\end{align}
for $i \in \lbrace 1, 2, \cdots, N \rbrace$ and for $k \in \lbrace 2, 3,  \cdots, N \rbrace.$ We refer to system~\eqref{eq:MOVM123} as the Modified Optimal Velocity Model (MOVM). Notice that, in contrast to~\eqref{eq:OVMO}, the MOVM no longer possesses the circular structure resulting from the periodic boundary condition. Indeed, the second vehicle (with index $1$) now follows the lead vehicle rather than the vehicle with index $N,$ since we consider the platoon to be traversing an infinitely long road.

The MOVM is described by a system of Delay Differential Equations (DDEs). Since such systems are hard to analyze, we obtain conditions for their local stability by analyzing them in the neighborhood of their equilibria. Such an analysis technique is called \emph{local stability analysis}. To obtain the equilibrium, we equate the $2N$ derivatives in~\eqref{eq:MOVM123} to zero. This yields $v_i^* = 0,$ $y_i^* = V^{-1}(\dot{x}_0),$ $i = 1, 2, \cdots, N$ as an equilibrium for system~\eqref{eq:MOVM123}. Therefore, to linearize~\eqref{eq:MOVM123} about this equilibrium, we first consider a small perturbation $u_i(t)$ about the equilibrium of the relative spacing pertaining to vehicle indexed $i.$ That is, $u_i(t)$ = $y_i(t)$ - $y_i^*.$ Next, we consider the Taylor's series expansion of $u_i(t)$, and set the leader's profile to zero, to obtain the linearized model, given by
%\begin{small}
\begin{align}
\nonumber
\dot{v}_1(t) = & \, - d u_1(t - \tau_1) - a v_1(t - \tau_1), \\ \nonumber
\dot{v}_k(t) = & \, d u_{k-1}(t - \tau_{k-1}) - d u_{k}(t - \tau_{k}) - a v_k(t - \tau_k), \\ \label{eq:LMOVM}
\dot{u}_i(t) = & \, v_i(t),
\end{align}
%\end{small}
for $i \in \lbrace 1, 2, \cdots, N \rbrace$ and for $k \in \lbrace 2, 3,  \cdots, N \rbrace$. Here, $d = a V^{'}(V^{-1}(\dot{x}_0))$ is the equilibrium coefficient, where the prime indicates differentiation with respect to a state variable. Henceforth, we denote $\tilde{d} = V^{'}(V^{-1}(\dot{x}_0)).$ Therefore, $d = a \tilde{d}.$

The MOVM is completely specified by the relative velocities $v_i$'s and the headways $y_i$'s. Therefore, the state of the MOVM at time ``$t$'' is given by $\textbf{S}(t)$ = $[ v_1(t) \text{ } v_2(t) \cdots v_N(t) \text{ } u_1(t) \text{ } u_2(t) \cdots u_N(t) ]^T \, \in \, \mathbb{R}^{2N} .$ Thus, system~\eqref{eq:LMOVM} can be succinctly written in matrix form as
\begin{align}
\label{eq:LMMOVM}
\dot{\textbf{S}}(t) = \sum\limits_{k = 0}^{N} A_k \textbf{S}(t - \tau_k),
\end{align}
where the matrices $A_k \in \mathbb{R}^{2N \times 2N}$ for each $k$. Also, $\tau_0$ is introduced for notational brevity and set to zero. The matrices $A_k, k = 1, 2 \cdots, N,$ are defined as follows.
\begin{align*}
A_0 =
\begin{bmatrix}
0_{N \times N} & 0_{N \times N} \\
I_{N \times N} & 0_{N \times N}
\end{bmatrix} .
\end{align*}
Here, $0_{N \times N}$ and $I_{N \times N}$ denote zero and identity matrices of order $N \times N$ respectively. For $1 \leq k \leq N-1$, we have
\begin{align*}
(A_k)_{ij} = \begin{cases}
- a , & i = j = k,\\
- d , & i = k , j = N + k,\\
 d , & i = k + 1, j = k, \\
0, & \text{elsewhere},
\end{cases}
\end{align*}
and
\begin{align*}
(A_N)_{ij} = \begin{cases}
- a , & i = j = N,\\
- d , & i = N , j = 2N,\\
0, & \text{elsewhere}.
\end{cases}
\end{align*}

\subsection{Optimal Velocity Functions (OVFs)}

There are several functions that satisfy the properties mentioned in Section~\ref{subsec:OVM}. We mention four widely-used OVFs~\cite{MBET}, obtained by fixing a functional form for $V(\cdot)$.
\begin{itemize}
\item[(a)] \emph{Underwood OVF:}
\begin{align*}
V_1(y) = V_0 e^{-\frac{2 y_m}{y}}.
\end{align*}
\item[(b)] \emph{Bando OVF:}
\begin{align*}
V_2(y) = V_0 \left( \tanh\left( \frac{y - y_m}{\tilde{y}} \right) + \tanh\left( \frac{y_m}{\tilde{y}} \right) \right).
\end{align*}
\item[(c)] \emph{Trigonometric OVF:}
\begin{align*}
V_3(y) = V_0 \left( \tan^{-1}\left( \frac{y - y_m}{\tilde{y}} \right) + \tan^{-1}\left( \frac{y_m}{\tilde{y}} \right) \right).
\end{align*}
\item[(d)] \emph{Hyperbolic OVF:}
\begin{align*}
V_4(y) = \begin{cases}
0, & y \leq y_0, \\
V_0 \left(\frac{(y - y_0)^n}{(\tilde{y})^n + (y - y_0)^n}\right), & y\geq y_0.
\end{cases}
\end{align*}
\end{itemize}
Here, $V_0,$ $y_0,$ $y_m,$ $\tilde{y}$ and $n$ are model parameters.

As captured by~\cite[Figure~$1$]{GKKITS}, the aforementioned OVFs behave similarly with varying headway. The following are noteworthy: $(i)$ The values attained by these OVFs, in the vicinity of the equilibrium, are almost the same, $(ii)$ their slopes, evaluated at the equilibrium, are different. The linearized version of the MOVM, captured by system~\eqref{eq:LMMOVM}, brings forth the dependence on the slope via the variable $\tilde{d},$ and $(iii)$ we make use of the Bando OVF throughout this paper, except in Section~\ref{section:RCCFMhopf}. Therein, we consider both the Bando OVF and the Underwood OVF, consistent with~\cite{GKK}.

We now proceed to understand the dynamical behavior of a platoon of cars running the MOVM.

%% file: sec3_ideal.tex
We first consider the idealistic case of instantaneously-reacting drivers. This results in zero reactions delays. Therefore, the model described by system~\eqref{eq:LMMOVM} boils down to the following system of Ordinary Differential Equations (ODEs):
\begin{align}
\label{eq:LMMOVMODE}
\dot{\textbf{S}}(t) = \, \left( \sum\limits_{k = 0}^{N} A_k \right) \textbf{S}(t).
\end{align}
We denote by $A,$ the sum of matrices $A_k,$ which is known as the \emph{dynamics matrix}. To characterize the stability of system~\eqref{eq:LMMOVMODE}, we require the eigenvalues of $A$ to be negative~\cite[Theorem 5.1.1]{GL}. To that end, we compute its characteristic function as
%\begin{align*}
%f(\lambda) & = \text{det}(\lambda I_{2N \times 2N} - A), \\
%& = \text{det}\left(\begin{bmatrix}
%(\lambda + a) I_{N \times N} & \hat{D} \\
%I_{N \times N} & \lambda I_{N \times N}
%\end{bmatrix} \right) = 0,
%\end{align*}
\begin{align*}
f(\lambda) = \text{det}(\lambda I_{2N \times 2N} - A) = \text{det}\left(\begin{bmatrix}
(\lambda + a) I_{N \times N} & \hat{D} \\
I_{N \times N} & \lambda I_{N \times N}
\end{bmatrix} \right) = 0,
\end{align*}
where $\hat{D}$ is derived from the dynamics matrix $A.$ The diagonal matrices of the above block matrix are invertible, and the off-diagonal matrices commute with each other. Hence, from~\cite[Theorem 3]{JS}, the characteristic equation can be simplified to $(\lambda^2 + a \lambda - d)^N = 0,$ which holds true if and only if
\begin{align}
\label{eq:LMOVMchareq}
\lambda^2 + a \lambda - d = 0.
\end{align}
Solving the above quadratic, we notice that the poles corresponding to system~\eqref{eq:LMMOVMODE} will be negative if $a > 0$ and $\tilde{d}  = V^{'}(V^{-1}(\dot{x}_0)) > 0.$ We note that, from physical constraints, $a > 0.$ Also, since $V(\cdot)$ is an OVF, it is monotonically increasing. Therefore, $ \tilde{d} > 0.$ Hence, for all physically relevant values of the parameters, the corresponding poles will lie in the open left-half of the Argand plane. Thus, the MOVM is locally stable for all physically relevant values of the parameters, in the absence of delays.

%% file: sec4_small.tex
Having studied the MOVM in the absence of reaction delays, we now analyze it in the small-delay regime. A way to obtain insight for the case of small delays is to conduct a linearization on time. This would yield a system of ODEs, which serves as an approximation to the original infinite-dimensional system~\eqref{eq:LMMOVM}, valid for small delays. We derive the criterion for such a system of ODEs to be stable, thereby emphasizing the design trade-off inherent among various system parameters and the reaction delay.

We begin by applying the Taylor's series approximation to the time-delayed state variables thus: $ v_i(t-\tau_i) \approx v_i(t) - \tau_i \dot{v}_i(t), $ and  $ u_i(t-\tau_i) \approx u_i(t) - \tau_i \dot{u}_i(t).$ Using this approximation for terms in~\eqref{eq:LMOVM}, substituting $v_i(t)$ for $\dot{u}_i(t)$ and re-arranging the resulting equations, we obtain the matrix equation
\begin{align}
\label{eq:LMOVMsmallmatrix}
B \dot{\textbf{S}}(t) = A \textbf{S}(t).
\end{align}
where the matrix $A$ is the dynamics matrix, as defined in Section~\ref{sec:ideal}, and $B$ is a block matrix of the form
\begin{align*}
B = 
\begin{bmatrix}
B_s & 0_{N \times N} \\
0_{N \times N} & I_{N \times N},
\end{bmatrix},
\end{align*}
where
\begin{align*}
(B_{s})_{ij} = \begin{cases}
1 - a \tau_i, & i = j, \\
0, & \text{elsewhere}.
\end{cases}
\end{align*}
We note that, since $B_s$ is a diagonal matrix, so is $B.$ Also, $B$ is invertible if and only if $a \tau_i \neq 1,$ for each $i.$ Thus, when $a \tau_i \neq 1,$ for each $i,$ we define $\tilde{C} = B^{-1}A,$ which is of the form
\begin{align*}
\tilde{C} = 
\begin{bmatrix}
\tilde{C}_s & \tilde{C}_c \\
I_{N \times N} & 0_{N \times N},
\end{bmatrix},
\end{align*}
where
\begin{align*}
(C_s)_{ij} = \begin{cases}
\frac{- a + d \tau_i}{1 - a \tau_i} , & i = j,\\
\frac{- d \tau_j}{1 - a \tau_i} , &  j = i - 1,\\
0, & \text{elsewhere},
\end{cases}
\end{align*}
and
\begin{align*}
(C_c)_{ij} = \begin{cases}
\frac{-  d}{1 - a \tau_i} , & i = j,\\
\frac{d}{1 - a \tau_i} , &  j = i - 1,\\
0, & \text{elsewhere}.
\end{cases}
\end{align*}

For system~\eqref{eq:LMOVMsmallmatrix} to be stable, the real part of eigenvalues of $\tilde{C}$ must be negative~\cite[Theorem 5.1.1]{GL}. To that end, we compute its characteristic function as
\begin{align*}
f(\lambda) = \text{det}(\lambda I_{2N \times 2N} - \tilde{C}) = \text{det}\left(\begin{bmatrix}
\lambda I_{N \times N} - \tilde{C}_s & - \tilde{C}_c \\
-I_{N \times N} & \lambda I_{N \times N}
\end{bmatrix} \right) = 0.
\end{align*}
%\begin{align*}
%f(\lambda) & = \text{det}(\lambda I_{2N \times 2N} - \tilde{C}), \\
%& = \text{det}\left(\begin{bmatrix}
%\lambda I_{N \times N} - \tilde{C}_s & - \tilde{C}_c \\
%-I_{N \times N} & \lambda I_{N \times N}
%\end{bmatrix} \right) = 0,
%\end{align*}
The diagonal matrices of the aforementioned block matrix are invertible, and the matrices in the second row therein commute with each other. Hence, the characteristic equation simplifies to~\cite[Theorem 3]{JS}
\begin{align*}
f(\lambda) = \text{det} \left( \lambda (\lambda I_{N \times N} - \tilde{C}_s) -  \tilde{C}_c \right) = 0.
\end{align*}
On further simplification, this yields
\begin{align*}
f(\lambda) = \overset{N}{\underset{i = 1}{\prod}} \left( (1 - a \tau_i) \lambda^2 + (a - d \tau_i) \lambda + d \right) = 0.
\end{align*}
For multiple terms in the above product to equal zero, their respective reaction delays must be equal. Such a possibility is not realistic, hence we ignore it. Therefore, for some $i \in \lbrace  1, 2, \cdots, N \rbrace,$ we have
\begin{align*}
(1 - a \tau_i) \lambda^2 + (a - d \tau_i) \lambda + d = 0.
\end{align*}
The roots of this quadratic equation are given by
\begin{align*}
\lambda_{1,2} = \frac{-(a-d\tau_i) \pm \sqrt{(a-d\tau_i)^2 - 4d(1-a\tau_i)}}{2(1-a\tau_i)}.
\end{align*}
We now consider the following (exhaustive) cases.
\begin{itemize}
\item[(1)] Let $a \tau_i > 1.$ Since $d > 0,$ it follows that $4d(1 - a\tau_i) < 0.$ Then, the eigenvalues are real. Further, one of these eigenvalues will be positive and the other negative. Hence, we require $a \tau_i <1$ for system~\eqref{eq:LMOVMsmallmatrix} to be stable.
\item[(2)] Let $(a-d\tau_i)^2 \geq 4d(1-a\tau_i).$ Then, the eigenvalues are real. They are negative if and only if $a-d\tau_i > 0,$ $i.e.,$ $\tilde{d} \tau_i < 1.$ Hence, we require $\tilde{d} \tau_i <1$ for system~\eqref{eq:LMOVMsmallmatrix} to be stable.
\item[(3)] Let $(a-d\tau_i)^2 < 4d(1-a\tau_i).$ Then, the eigenvalues are complex. The real part of the eigenvalues will be negative if and only if $a-d\tau_i > 0,$ $i.e.,$ $\tilde{d} \tau_i < 1.$ Hence, we require $\tilde{d} \tau_i <1$ for system~\eqref{eq:LMOVMsmallmatrix} to be stable.
\end{itemize}
From the above cases, it is clear that system~\eqref{eq:LMOVMsmallmatrix} is stable if and only if
\begin{align}
\label{eq:LMOVMsmallcond}
\max(a,\tilde{d}) \tau_i <1,
\end{align}
for each $i \in \lbrace 1, 2, \cdots, N \rbrace.$ Recall that we obtained system~\eqref{eq:LMOVMsmallmatrix} by truncating the Taylor's series to first order. Hence,~\eqref{eq:LMOVMsmallcond} is a sufficient condition for the local stability of the MOVM described by system~\eqref{eq:MOVM123}, valid for small values of the reaction delay.

%We note that this case is motivated by the futuristic self-driven vehicles, wherein the delays are expected to be small~\cite{AK}. From a given control algorithm, the OVF is fixed. Hence, $\tilde{d}$ may not be varied by the control algorithm. The working of the control algorithm is as follows. First, the reaction delay is estimated for each vehicle, and the largest among them chosen. Next, an appropriate sensitivity parameter $a$ is chosen that ensures stability.

%% file: sec5_hopf.tex
Having studied system~\eqref{eq:MOVM123} in the no-delay and the small-delay regimes, in this section, we focus on the arbitrary-delay regime. We first derive the necessary and sufficient condition for the local stability of the MOVM. We then show that, upon violation of this condition, the corresponding traffic flow transits via a Hopf bifurcation to the locally unstable regime.

\subsection{Transversality condition}
\label{section:MOVMTC}

Hopf bifurcation is a phenomenon wherein, on appropriate variation of system parameters, a dynamical system either loses or regains stability because of a pair of conjugate eigenvalues crossing the imaginary axis in the Argand plane~\cite[Chapter 11, Theorem 1.1]{HL}. Mathematically, Hopf bifurcation analysis is a rigorous way of proving the emergence of limit cycles (isolated closed trajectory in state space) in non-linear dynamical systems.

To determine if system~\eqref{eq:MOVM123} undergoes a stability loss via a Hopf bifurcation, we follow~\cite{GR} and introduce an exogenous, non-dimensional parameter $\kappa > 0.$ A general system of DDEs
\begin{align}
\label{eq:origDDE}
\dot{x}(t) = f(x(t), x(t-\tau_1), \cdots, x(t-\tau_n)),
\end{align}
is modified to
\begin{align}
\label{eq:modDDE}
\dot{x}(t) = \kappa f(x(t), x(t-\tau_1), \cdots, x(t-\tau_n)),
\end{align}
with the introduction of the exogenous parameter. Note that $(i)$ $\kappa$ has no effect on the equilibrium of system~\eqref{eq:origDDE}, and $(ii)$ we obtain system~\eqref{eq:origDDE} by setting $\kappa = 1$ in system~\eqref{eq:modDDE}. We first linearize system~\eqref{eq:modDDE} about its non-trivial equilibrium and derive its characteristic equation. We then search for a pair of conjugate eigenvalues on the imaginary axis in the Argand plane. This yields the necessary and sufficient condition for the local stability of system~\eqref{eq:modDDE}. Setting the exogenous parameter to unity then yields the necessary and sufficient condition for system~\eqref{eq:origDDE}. The exogenous parameter so introduced helps simplify the requisite algebra and capture any interdependence among the system parameters.

For the MOVM, introducing $\kappa$ in~\eqref{eq:MOVM123} yields
\begin{align}
\nonumber
\dot{v}_1(t) = & \,  \ddot{x}_0(t) + \kappa a \left( \dot{x}_0(t - \tau_1) -  V(y_1(t - \tau_1)) - v_1(t - \tau_1) \right) , \\ \nonumber
\dot{v}_k(t) = & \, \kappa a \left( V(y_{k-1}(t - \tau_{k-1})) - V(y_{k}(t - \tau_k)) - v_k(t - \tau_k) \right), \\ \label{eq:modMOVM123eqn}
\dot{y}_i(t) = & \, \kappa v_i(t),
\end{align}
for $i \in \lbrace 1, 2, \cdots, N \rbrace$ and for $k \in \lbrace 2, 3,  \cdots, N \rbrace.$ We linearize this about the equilibrium $v_i^* = 0,$ $y_i^* = V^{'}(V^{-1}(\dot{x}_0)),$ $i = 1, 2, \cdots, N,$ and write it in matrix form to obtain
\begin{align}
\label{eq:modLMMOVMeqn}
\dot{\textbf{S}}(t) = \sum\limits_{k = 0}^{N} \tilde{A}_k \textbf{S}(t - \tau_k),
\end{align}
where the matrices $\tilde{A}_k = \kappa A_k,$ for $k = 1, 2, \cdots, N,$ where the matrices $A_k$ are as defined in Section~\ref{sec:models}.

The characteristic equation corresponding to system~\eqref{eq:modLMMOVMeqn} is obtained as~\cite[Section 5.1]{GL}
 $$f(\lambda) = \text{det} \left(\lambda I_{2N \times 2N} - \sum\limits_{k = 0}^{N} e^{- \lambda \tau_k} \tilde{A}_k \right) = 0.$$
The matrix in consideration is a block matrix of the form
\begin{align*}
\lambda I_{2N \times 2N} - \sum\limits_{k = 0}^{N} e^{- \lambda \tau_k} \tilde{A}_k =
\begin{bmatrix}
\tilde{A} & \tilde{B} \\
\tilde{C} & \tilde{D}
\end{bmatrix} ,
\end{align*}
where $\tilde{C} = -\kappa I_{N \times N}$ and $\tilde{D} = \lambda I_{N \times N}.$ Further, $\tilde{A}$ is a diagonal matrix with the $i^{th}$ diagonal entry being $\lambda + \kappa a e^{-\lambda \tau_i},$ and $\tilde{B}$ is a sparse lower-triangular matrix. Clearly, $\tilde{A}$ and $\tilde{D}$ are invertible, and $\tilde{C}$ commutes with $\tilde{D}$. Therefore, the characteristic equation simplifies to the form~\cite[Theorem 3]{JS}
\begin{align*}
f(\lambda) = \text{det} \left( \begin{bmatrix}
\tilde{A} & \tilde{B} \\
\tilde{C} & \tilde{D}
\end{bmatrix}  \right) = \text{det} \left( \tilde{A} \tilde{D} - \tilde{B} \tilde{C} \right) = 0.
\end{align*}
Simplifying the above expression, we obtain the characteristic equation pertaining to~\eqref{eq:modLMMOVMeqn} as
\begin{align}
\label{eq:CELMMOVM}
f(\lambda) = \overset{N}{\underset{i = 1}{\prod}} (\lambda^2 + \kappa a \lambda e^{- \lambda \tau_i} + \kappa^2 d e^{- \lambda \tau_i}) = 0.
\end{align}
For multiple terms in the above product to equal zero, their respective reaction delays must be equal. Such a possibility is not realistic, hence we ignore it. Therefore, for some $i \in \lbrace  1, 2, \cdots, N \rbrace,$ we have
\begin{align}
\label{eq:CELMMOVMK}
\lambda^2 + \kappa a \lambda e^{- \lambda \tau_i} + \kappa^2 d e^{- \lambda \tau_i} = 0.
\end{align}

System~\eqref{eq:modMOVM123eqn} will be locally stable if and only if all the roots of~\eqref{eq:CELMMOVMK} lie in the open left-half of the Argand plane~\cite[Theorem 5.1.1]{GL}. Therefore, we search for a conjugate pair of eigenvalues of~\eqref{eq:CELMMOVMK} that crosses the imaginary axis in the Argand plane. To that end, we substitute $\lambda = j \omega$ in~\eqref{eq:CELMMOVMK}, with $j = \sqrt{-1}.$ We then equate the real and imaginary parts to zero and obtain
\begin{align}
\label{eq:simeqnMOVM1}
\kappa a \omega \text{ sin}(\omega \tau_i) +  \kappa^2 d \text{ cos}(\omega \tau_i) = & \, \omega^2, \\ \label{eq:simeqnMOVM2}
\kappa a \omega \text{ cos}(\omega \tau_i) -  \kappa^2 d \text{ sin}(\omega \tau_i) = & \, 0.
\end{align}
Squaring and adding~\eqref{eq:simeqnMOVM1} and~\eqref{eq:simeqnMOVM2} yields $\omega^4 - \kappa^2 a^2 \omega^2 - \kappa^4 d^2 = 0.$ Solving for $\omega^2$, we obtain
\begin{align*}
\omega_{1,2}^2 = \kappa^2 \left( \frac{a^2 \pm \sqrt{a^4 + 4 d^2}}{2} \right).
\end{align*}
Since we are searching for a positive root, we discard the negative root. The positive root of the above expression is given by
\begin{align}
\omega = \kappa \sqrt{\frac{a(a + \sqrt{a^2 + 4 \tilde{d}^2})}{2}}.
\end{align}
For convenience, we write the above equation as $\omega = \kappa \chi.$ Notice that, on re-arranging~\eqref{eq:simeqnMOVM2}, we obtain $\kappa \tilde{d} \tan(\omega \tau_i) = \omega.$ Substituting for $\omega$ in the above equation and simplifying yields
\begin{align}
\label{eq:omkaMOVM1}
\omega_0 = \frac{1}{\tau_i} \tan^{-1} \left( \frac{\chi}{\tilde{d}} \right).
\end{align}
Substituting $\omega_0$ in~\eqref{eq:simeqnMOVM2} and simplifying, we obtain
\begin{align}
\label{eq:omkaMOVM2}
\kappa_{cr}  = \frac{1}{\tau_i \chi} \tan^{-1} \left( \frac{\chi}{\tilde{d}} \right) .
\end{align}
Thus,~\eqref{eq:omkaMOVM1} and~\eqref{eq:omkaMOVM2} yield the angular frequency of the oscillatory solution and the value of $\kappa$ at which such a solution exists respectively.

We now show that the MOVM undergoes a Hopf bifurcation at $\kappa = \kappa_{cr}.$ To that end, we need to prove the transversality condition of the Hopf spectrum. That is, we must show that~\cite[Chapter 11, Theorem 1.1]{HL}
\begin{align}
\label{eq:HopfTransCond}
\text{Real}\left( \frac{\text{d} \lambda}{\text{d} \kappa} \right)_{\kappa = \kappa_{cr} } \neq \, 0.
\end{align}
To that end, we differentiate~\eqref{eq:CELMMOVMK} with respect to $\kappa$ and simplify it, to obtain
\begin{align}
\label{eq:MOVMtranscond}
\text{Real}\left( \left( \frac{\text{d} \lambda}{\text{d} \kappa} \right)^{-1} \right)_{\kappa = \kappa_{cr} } = \frac{\kappa_{cr} \omega_0^2 \tau_i(\kappa_{cr}^2 \tilde{d} \cos(\omega_0 \tau_i) + \omega_0^2)}{(\kappa_{cr}^2 \tilde{d} \cos(\omega_0 \tau_i) + \omega_0)^2 + (\kappa_{cr}^2 \tilde{d} \sin(\omega_0 \tau_i))^2} > 0.
\end{align}
The positivity in~\eqref{eq:MOVMtranscond} follows because $\cos(\omega_0 \tau_i)$ $=$ $\kappa_{cr} \tilde{d}/(\kappa_{cr}^2 \tilde{d}^2 + \omega_0^2)$ is positive. This expression follows from~\eqref{eq:simeqnMOVM2} using trigonometric manipulations. Also, Real$(z) > 0$ if and only if Real$\left(1/z\right) > 0$ $\forall$ $z \in \mathbb{C}.$ Hence, from~\eqref{eq:MOVMtranscond} we have
\begin{align*}
\text{Real}\left( \frac{\text{d} \lambda}{\text{d} \kappa} \right)_{\kappa = \kappa_{cr} } > 0.
\end{align*}
This proves the transversality of the Hopf spectrum. Therefore, the MOVM transits from the locally stable to the locally unstable regime via a Hopf bifurcation at $\kappa = \kappa_{cr}.$ It can be shown that for sufficiently small values of $\kappa,$~system~\eqref{eq:modMOVM123eqn} is locally stable. Additionally, the above strict inequality implies that the eigenvalues move from left to right in the Argand plane as $\kappa$ is increased in the neighborhood of $\kappa_{cr}.$ Therefore, $\kappa < \kappa_{cr}$ is the necessary and sufficient condition for local stability of system~\eqref{eq:modMOVM123eqn}.

\subsection{Discussion}
\label{section:MOVMdisc}
A few comments are in order.
\begin{itemize}
\item[(1)] Note that the characteristic equation~\eqref{eq:CELMMOVMK} is transcendental, hence there exist infinitely many roots. However, system~\eqref{eq:modMOVM123eqn} loses local stability when the first conjugate pair of eigenvalues crosses the imaginary axis as the exogenous parameter is varied. Due to the positivity of the derivative in~\eqref{eq:MOVMtranscond}, system stability cannot be restored by increasing $\kappa.$
\item[(2)] The equation of the stability boundary pertaining to system~\eqref{eq:modMOVM123eqn} is $\kappa = \kappa_{cr}.$ It is also called the Hopf boundary of the said system. To obtain the Hopf boundary corresponding to the MOVM described by system~\eqref{eq:MOVM123}, we tune the system parameters such that $\kappa_{cr} = 1$ in~\eqref{eq:omkaMOVM2}. In particular, the MOVM is locally stable if and only if, for each $i \in \lbrace 1, 2, \cdots, N \rbrace,$ we have
\begin{align}
\label{eq:nscMOVM}
 \tau_i < \frac{1}{\chi} \tan^{-1} \left( \frac{\chi}{\tilde{d}} \right).
\end{align}
\begin{figure}[t]
\centering
\includegraphics[scale=0.26,angle=90]{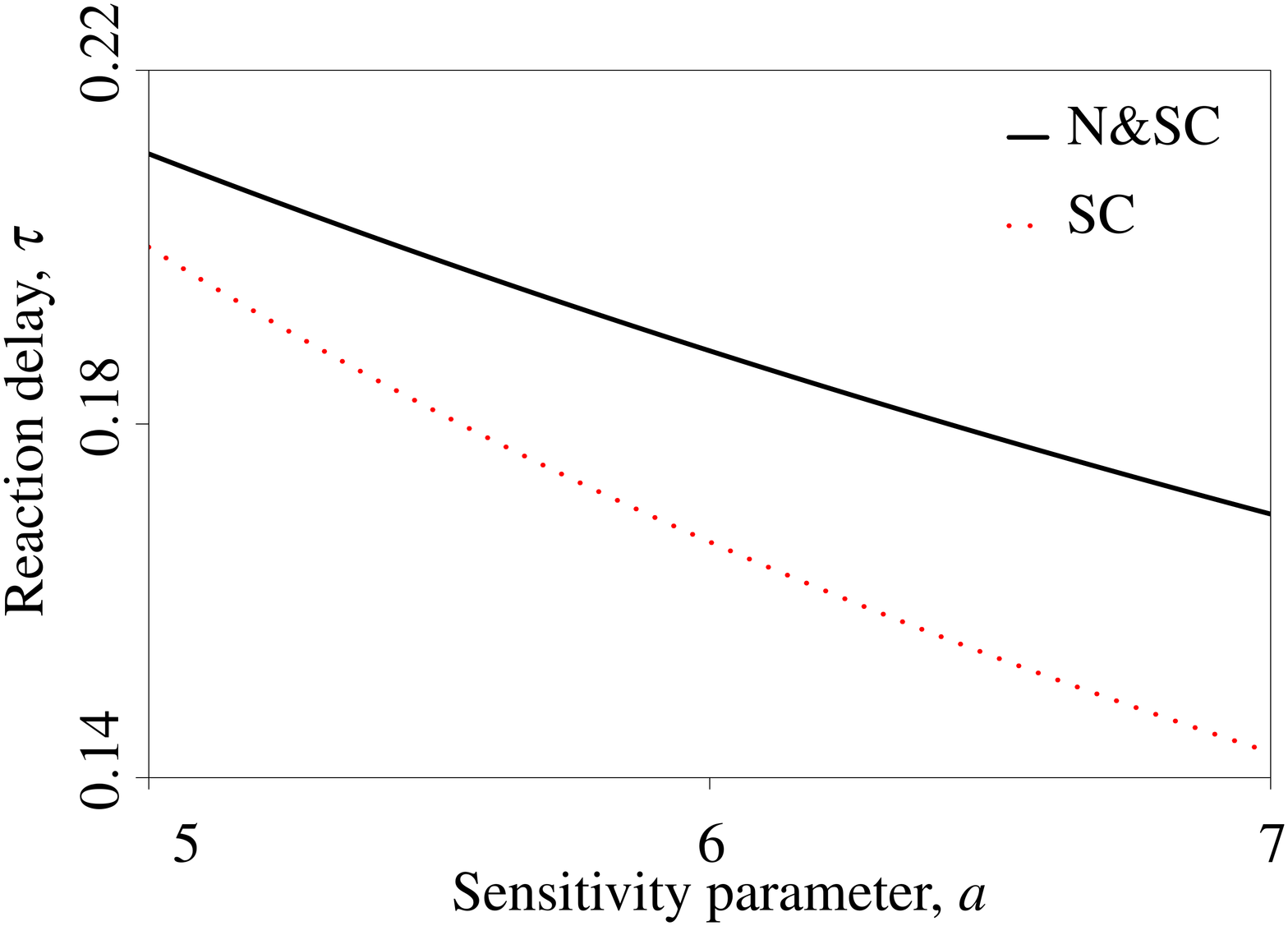}
\caption{\emph{Stability chart:} Illustrates  the necessary and sufficient condition (N\&SC)~\eqref{eq:nscMOVM} and the sufficient condition (SC)~\eqref{eq:LMOVMsmallcond} for the MOVM, for small delays. The plot serves to validate our analysis presented in Section~\ref{sec:small}.}\label{fig:stab_chart}
\end{figure}
It is clear from~\eqref{eq:nscMOVM} that when the reaction delay increases, the MOVM loses local stability via a Hopf bifurcation. Also note that when $\tau = 0,$~\eqref{eq:nscMOVM} is trivially satisfied for all physically relevant parameter values. This is in agreement with the result derived in Section~\ref{sec:ideal}. To validate the analysis presented in Section~\ref{sec:small}, we plot the Right Hand Sides (RHSs) of~\eqref{eq:LMOVMsmallcond} and~\eqref{eq:nscMOVM} for small values of the reaction delay in Fig.~\ref{fig:stab_chart}. Clearly, we notice from Fig.~\ref{fig:stab_chart} that~\eqref{eq:LMOVMsmallcond} indeed represents a sufficient condition for the local stability of the MOVM for small delays.
\item[(3)] Loss of local stability via a Hopf bifurcation results in the emergence of limit cycles. Since the dynamical variables for the MOVM correspond to relative velocities and headways, these non-linear oscillations physically manifest as back-propagating congestion wave on a highway. Thus, as mentioned in the Introduction, our analysis provides a mathematical basis to the commonly-observed ``phantom jam.''
\item[(4)] Note that the non-dimensional parameter $\kappa$ is not a model parameter; rather, it is an exogenous mathematical entity introduced to aid the analysis and capture any interdependence among model parameters. It also serves to simplify the algebra required to obtain the necessary and sufficient condition for local stability of the MOVM. Further, since substituting $\kappa = 1$ yields the MOVM, it is useful in a neighborhood around $1,$ \emph{i.e.,} near the stability boundary.
\item[(5)] Gain parameters are known to destabilize feedback systems~\cite[Section 3.7]{RR}. Thus, we need to verify that the bifurcation phenomenon proved in this section is not an artefact of the exogenous parameter. To that end, we need to verify that the MOVM also undergoes a Hopf bifurcation when one of the model parameters is chosen as the bifurcation parameter. It was shown in~\cite{SM} that the transversality condition of the Hopf spectrum holds true for the characteristic equation of the form~\eqref{eq:CELMMOVMK} (with $\kappa = 1$) when $\tau$ is used as the bifurcation parameter, although in a different context.
\end{itemize}

%% file: sec6_noc.tex
In the previous three sections, we derived conditions for the MOVM to be locally stable in three different regimes. In the next two sections, we explore two important properties of the MOVM; namely, non-oscillatory convergence and the rate of convergence.

%In this section, we derive the necessary and sufficient condition for non-oscillatory convergence of the MOVM. Mathematically, this amounts to ensuring that the eigenvalues corresponding to system~\eqref{eq:MOVM123} are negative real numbers. Qualitatively, non-oscillatory convergence avoids jerky vehicular motion since relative velocities and headways constitute dynamical variables. Such results could help ensure the smooth flow of traffic, and hence improve the ride quality.

In the context of transportation networks, ride quality is of utmost importance. This, in turn, mandates that the vehicles avoid jerky motion. Since relative velocities and headways constitute dynamical variables for the MOVM, it boils down to studying the non-oscillatory property of its solutions. In particular, we derive the necessary and sufficient condition for non-oscillatory convergence of the MOVM. Mathematically, this amounts to ensuring that the eigenvalues corresponding to system~\eqref{eq:LMMOVM} are negative real numbers.

To derive the necessary and sufficient condition for non-oscillatory convergence of the MOVM, we begin with the characteristic equation corresponding to system~\eqref{eq:MOVM123}, obtained by setting $\kappa = 1$ in~\eqref{eq:CELMMOVMK}. We also drop the subscript ``$i$'' for convenience. Thus, we obtain
\begin{align}
\label{eq:RCELMMOVM}
f(\lambda) = \lambda^2 + (a \lambda + d ) e^{- \lambda \tau} = 0.
\end{align}

To ensure non-oscillatory convergence of the MOVM, we require the roots of~\eqref{eq:RCELMMOVM} to be real and negative. To that end, we substitute $\lambda = \sigma + j \omega$ in~\eqref{eq:RCELMMOVM}, where $j = \sqrt{-1}.$ This yields
\begin{align}
\label{eq:reeqzero}
a \omega \sin(\omega \tau) + (a \sigma + d) \cos(\omega \tau) & = (\omega^2 - \sigma^2) e^{\sigma \tau}, \text{ and } \\ \label{eq:imeqzero}
a \omega \sin(\omega \tau) + (a \sigma + d) \cos(\omega \tau) & = (- 2 \sigma \omega) e^{\sigma \tau}.
\end{align}
Squaring and adding~\eqref{eq:reeqzero} and~\eqref{eq:imeqzero}, we obtain
\begin{align}
\label{eq:halfcond}
(a \omega)^2 + (a \sigma + d)^2 = (\omega^2 + \sigma^2)^2 e^{2\sigma \tau}.
\end{align}
To ensure that the roots are real, we require a condition for $\omega = 0$ to be the only solution of~\eqref{eq:halfcond}. Substituting $\omega = 0$ in~\eqref{eq:halfcond}, we obtain
\begin{align}
\label{eq:necnoc}
(a \sigma + d)^2 = \sigma^4 e^{2\sigma \tau}.
\end{align}
Thus, the above condition is necessary for $\omega = 0$ to be a solution of~\eqref{eq:halfcond}. To check whether it is also a sufficient condition, we first separate the terms containing $\omega$ in~\eqref{eq:halfcond} from those without it. This yields
\begin{align*}
 e^{2\sigma \tau} \omega^4 + (2 \sigma^2  e^{2\sigma \tau} - a^2) \omega^2 = (a \sigma + d)^2 - \sigma^4 e^{2\sigma \tau}.
\end{align*}
Assuming $(a \sigma + d)^2 = \sigma^4 e^{2\sigma \tau},$ we solve the above quadratic in $\omega^2$ to obtain
\begin{align*}
\omega^2 = 0 \text{ or } \omega^2 = \frac{a^2 - 2 \sigma^2  e^{2\sigma \tau}}{e^{2\sigma \tau}}.
\end{align*}
Thus, for $\omega = 0$ to be the unique solution of~\eqref{eq:halfcond}, we require $a^2 = 2 \sigma^2  e^{2\sigma \tau}$ in addition to the condition mentioned in~\eqref{eq:necnoc}. That is,~\eqref{eq:RCELMMOVM} has real eigenvalues if and only if
\begin{align}
\label{eq:partialnscroc}
(a \sigma + d)^2 = \sigma^4 e^{2\sigma \tau}, \text{ and } a^2 = 2 \sigma^2  e^{2\sigma \tau}.
\end{align}
Solving the above two equations for the eigenvalue, we obtain
\begin{align}
\label{eq:realeig}
\sigma =  \tilde{d} m_{\pm}, \text{ with } m_{\pm} = -2 \pm \sqrt{2}. 
\end{align}
Notice from the foregoing analysis that the eigenvalues are guaranteed to be negative if they are real. Substituting~\eqref{eq:realeig} in~\eqref{eq:RCELMMOVM} and re-arranging, we obtain the boundary for the region of non-oscillatory convergence as
\begin{align*}
e^{-\tilde{d} \tau m_{\pm}} = \frac{- m_{\pm}^2 \tilde{d}}{a (m+1)}.
\end{align*}
Notice that the Left Hand Side (LHS) in the above equation is a non-negative quantity. The RHS is non-negative for $m_{-}$ but not for $m_{+}.$ Hence, we set $m = m_{-}$ in the above equation, and re-arrange to obtain
\begin{align}
\label{eq:boundaryMOVM}
\tau_{noc} = \frac{1}{m \tilde{d}} \text{ ln}\left(\frac{-a(m+1)}{m^2 \tilde{d}}\right),
\end{align}
where $\tau_{noc}$ represents the boundary for the region of non-oscillatory convergence in the $\tau$-domain. Therefore, $\tau < \tau_{noc}$ represents the necessary and sufficient condition for non-oscillatory convergence of the MOVM. We note that the following inequalities must be satisfied: $0 < \tau_{noc} < \tau_{cr},$ where $\tau_{cr}$ is the RHS of~\eqref{eq:nscMOVM}. 

In summary, the necessary and sufficient condition for non-oscillatory convergence of the MOVM is
\begin{align}
\label{eq:nscnoc}
\tau_{i} < \frac{1}{m \tilde{d}} \text{ ln}\left(\frac{-a(m+1)}{m^2 \tilde{d}}\right),
\end{align}
for each $i \in \lbrace 1, 2, \cdots, N \rbrace,$ when the RHS is positive and less than $\tau_{cr}.$

\begin{figure}[t]
\centering
\includegraphics[scale=0.26,angle=90]{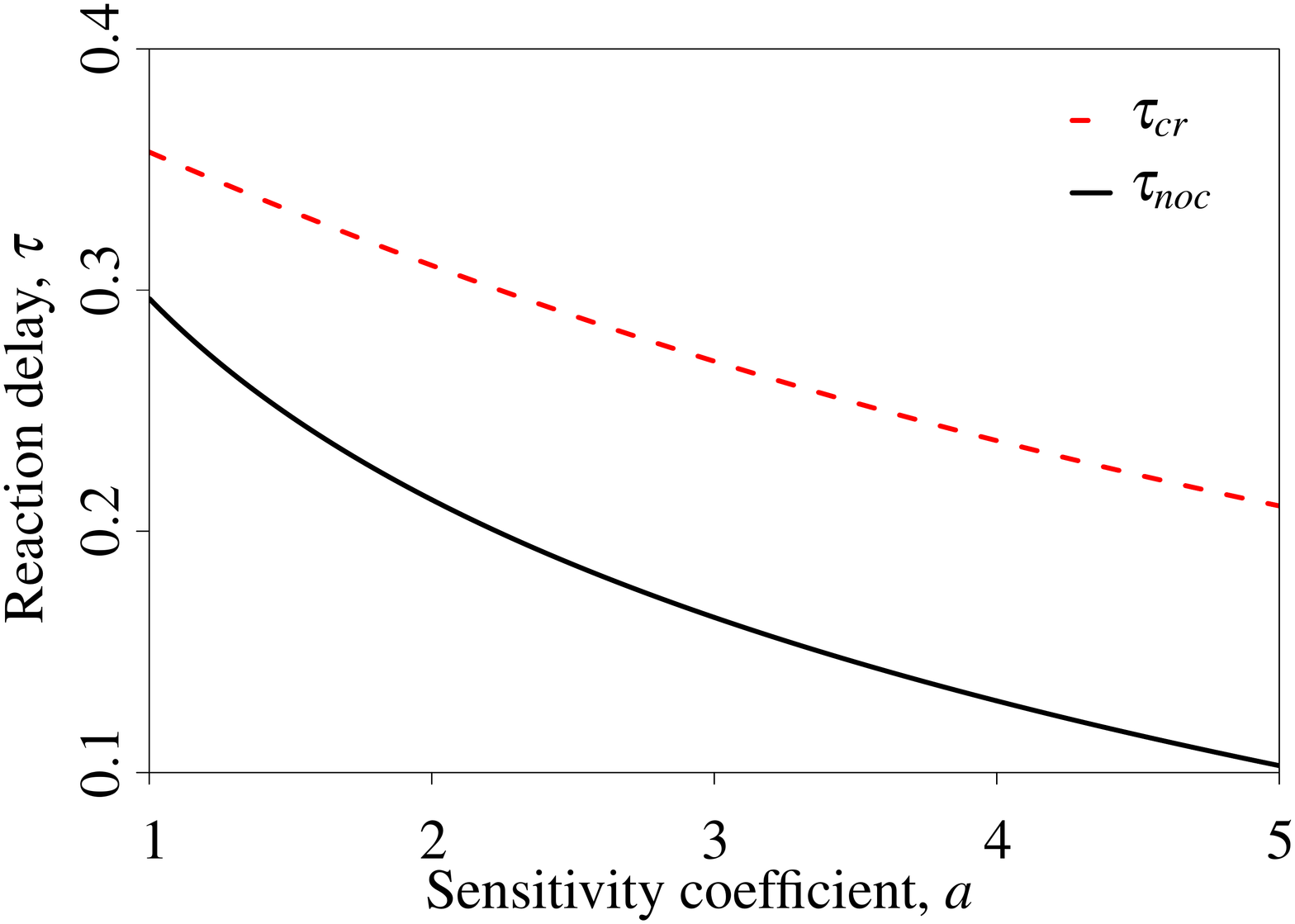}
\caption{Illustration of region of non-oscillatory convergence for the MOVM. Here, $\tau_{cr}$ and $\tau_{noc}$ represent the boundary of the locally stable region and the region of non-oscillatory convergence of the MOVM respectively. Notice the stringent requirements on the reaction delay for the solutions of the MOVM to be non-oscillatory.}\label{fig:nocbound}
\end{figure}

We now illustrate the boundary for the region of non-oscillatory convergence of the MOVM described by~\eqref{eq:boundaryMOVM}. In order to better-understand the stringent constraints on system parameters to achieve non-oscillatory convergence, we also plot the necessary and sufficient condition for local stability~\eqref{eq:nscMOVM} of the MOVM. To that end, we make use of the Bando OVF. We let the equilibrium velocity of the lead vehicle to be $\dot{x}_0 = 5$ m/s, and the model parameters as $y^* = 2$ m, $\tilde{y} = 5$ m and $y_m = 1$ m. We then compute the corresponding $V_0$ and $\tilde{d}.$ We vary the sensitivity coefficient $a$ from $1$ and $5,$ and compute the requisite boundaries using the scientific computation software MATLAB.

Fig.~\ref{fig:nocbound} portrays regions of local stability and non-oscillatory convergence for the MOVM in the $(a,\tau)$-space. For a fixed $a,$ the reaction delay must not exceed $\tau_{cr}$ (respectively, $\tau_{noc}$) for the MOVM to be locally stable (respectively, possess non-oscillatory solutions). Clearly, the values of $\tau$ need to be much smaller for the solutions of the MOVM to be non oscillatory as opposed to the stability of the MOVM, for a fixed value of $a.$ In fact, as the sensitivity parameter $a$ increases, the corresponding value of reaction delays required to ensure non-oscillatory convergence decreases rapidly.
%Fig.~\ref{fig:nonoscil} captures an instance of the velocity and the headway solution of vehicle $2$ in a $4$-vehicle platoon running the Hyperbolic OVF~\cite{GKKITS} with the following parameter values: $N = 3,$ $y_i^* = 30$~m, $i = 1,2,3,$ $a = 1.2$~$\text{s}^{-1},$ $\tilde{y} = 9$~m, $\tau_1 = 0.2$~s, $\tau_2 = \tau_3 = 0.01$~s. The parameter $V_0$ is computed under the influence of a leader whose equilibrium velocity is $25$~m/s. This instance captures the non-oscillatory convergence of state variables to their respective equilibria.

We end this section with two remarks. $(i)$ To the best of our knowledge, the analysis presented in this section is the first to address non-oscillatory convergence of systems with characteristic equations of the form~\eqref{eq:RCELMMOVM} using spectral-domain techniques. $(ii)$ We can obtain $\omega = 0$ as the only solution to~\eqref{eq:halfcond} by a geometrical method as follows. Re-arranging~\eqref{eq:halfcond} yields
\begin{align*}
(\omega^2 + \sigma^2)^2 = (a^2 e^{-2\sigma \tau}) \omega^2 + ( a\sigma + d )^2 e^{-2\sigma \tau}.
\end{align*}
Notice that the LHS and RHS of the above equation represent a parabola and a line in $\omega^2$-domain respectively. Since a parabola is strictly convex, the tangent to a parabola at any point will intersect it only at that point. Therefore, the slope of the line (captured by the coefficient of the term $\omega^2$) must equal the derivative of the LHS with respect to $\omega^2,$ evaluated at zero. That is, $a^2 e^{-2\sigma \tau} = 2\sigma^2.$ Also, the intercept of the line should be equal the value of the parabola evaluated at zero, \emph{i.e.,} $( a\sigma + d )^2 e^{-2\sigma \tau} = \sigma^4.$ These are same as the conditions in~\eqref{eq:partialnscroc}.
%Finally, we provide an alternate geometrical method to arrive at the conditions required for $\omega = 0$ to be the unique solution of~\eqref{eq:halfcond}. The said equation can be re-written as
%\begin{align*}
%(\omega^2 + \sigma^2)^2 = (a^2 e^{-2\sigma \tau}) \omega^2 + ( a\sigma + d )^2 e^{-2\sigma \tau}.
%\end{align*}
%The left hand side and right hand side of the above equation represent a parabola and a line in the variable $\omega^2,$ respectively. We wish for these to intersect only at $\omega^2 = 0.$ Since a parabola is convex, the tangent to a parabola at any given point intersects it only at that point. Hence, we wish for the right hand side to be the tangent at $\omega^2 = 0.$ To that end, we differentiate the left hand side of~\eqref{eq:halfcondalt} with respect to $\omega^2,$ and set $\omega^2 = 0$ in the resulting expression. This must equal $a^2 e^{-2\sigma \tau},$ thus yielding $a^2 e^{-2\sigma \tau} = 2 \sigma^2.$ The intercept of the right hand side of~\eqref{eq:halfcondalt} must equal the left hand side evaluated at zero. This yields $( a\sigma + d )^2 e^{-2\sigma \tau} = \sigma^4.$ These are precisely the conditions in~\eqref{eq:partialnscroc}.
%\begin{figure}[t]
%\centering
%\includegraphics[scale=0.25,angle=90]{solnon.pdf}
%\caption{Illustration of a non-oscillatory solution of the MOVM.}\label{fig:nonoscil}
%\end{figure}

%% file: sec7_roc.tex
In this section, we characterize the time required to attain the uniform traffic flow, once the traffic flow is perturbed (by events such as the departure of a vehicle from the platoon). Mathematically, it is related to the rate of convergence of solutions of the MOVM to the desired equilibrium. To that end, we follow~\cite{SC} and first characterize the rate of convergence of the MOVM. Then, using the notion of settling time, we derive an expression for the time a platoon takes to attain the desired equilibrium following a perturbation.

%Rate of convergence is an important performance metric that dictates the time a dynamical system takes to equilibrate, when perturbed. In the context of a transportation network, it is related to the time required to attain the uniform traffic flow, once the traffic flow is perturbed (by events such as the departure of a vehicle from the platoon). Following~\cite{SC}, we now characterize the rate of convergence for the MOVM.

We begin by recalling the characteristic equation pertaining to system~\eqref{eq:LMMOVM} from Section~\ref{section:MOVMTC}. Dropping the subscript ``$i$'' for ease of exposition, and setting $\kappa$ $=$ $1$ in~\eqref{eq:CELMMOVMK}, we obtain
\begin{align*}
\lambda^2 + a \lambda e^{- \lambda \tau} + d e^{- \lambda \tau} = 0.
\end{align*}
Using the change of variables $z = \lambda \tau,$ the above equation results in
\begin{align}
\label{eq:ROCMOVMT}
z^2 e^z + a^{*} z + d^{*} = 0,
\end{align}
where $a^{*} = a \tau$ and $d^{*} = d \tau^2.$ Notice that~\eqref{eq:ROCMOVMT} has the same form as~\cite[Equation $(22)$]{SC}. Hence, following~\cite{SC}, we substitute $z$ $=$ $\psi - \sigma,$ where $\sigma$ is non-negative and real, in~\eqref{eq:ROCMOVMT} to obtain
\begin{align*}
(\psi^2 - 2 \sigma \psi + \sigma^2) e^{\psi} + a^* e^{\sigma} \psi + (d^* - a^* \sigma)e^{\sigma} = 0.
\end{align*} 

%\begin{figure}[t]
%\centering
%\includegraphics[scale=0.25,angle=0]{roccnt.pdf}
%\caption{Contour lines of the rate of convergence overlaying the boundaries of the locally stable region and the region of non-oscillatory convergence. Notice the trade-off between the rate of convergence and non-oscillatory convergence; very high rates of convergence cannot be achieced if the solutions are to posses the non-oscillatory property.}\label{fig:roccnt}
%\end{figure}

\begin{figure*}[t]
 \begin{center}
 \subfloat[]{
 \includegraphics[scale=0.26,angle=90]{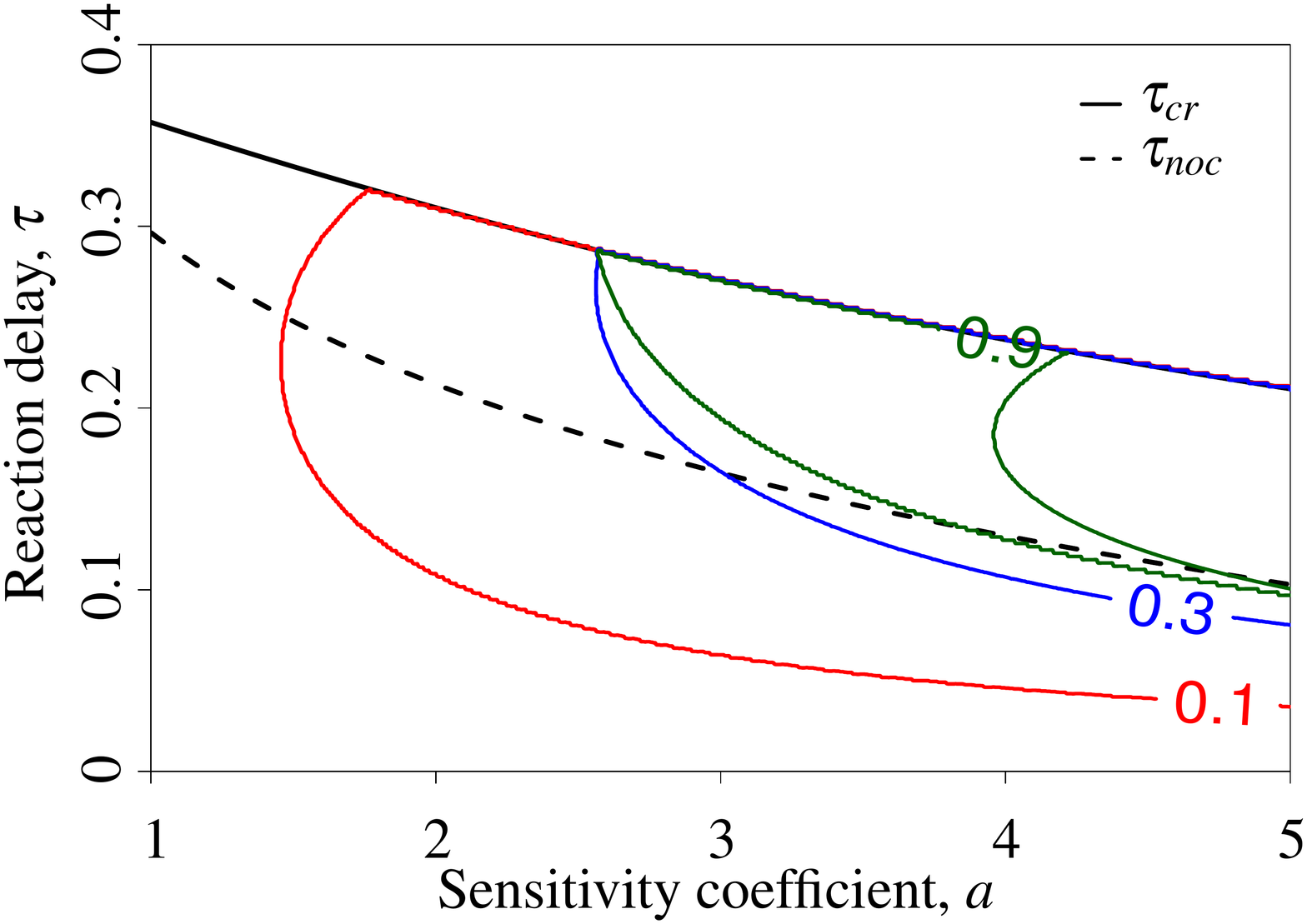}
  \label{fig:roccnt1} 
  } \hspace{1mm}
  \subfloat[]{
\includegraphics[scale=0.26,angle=90]{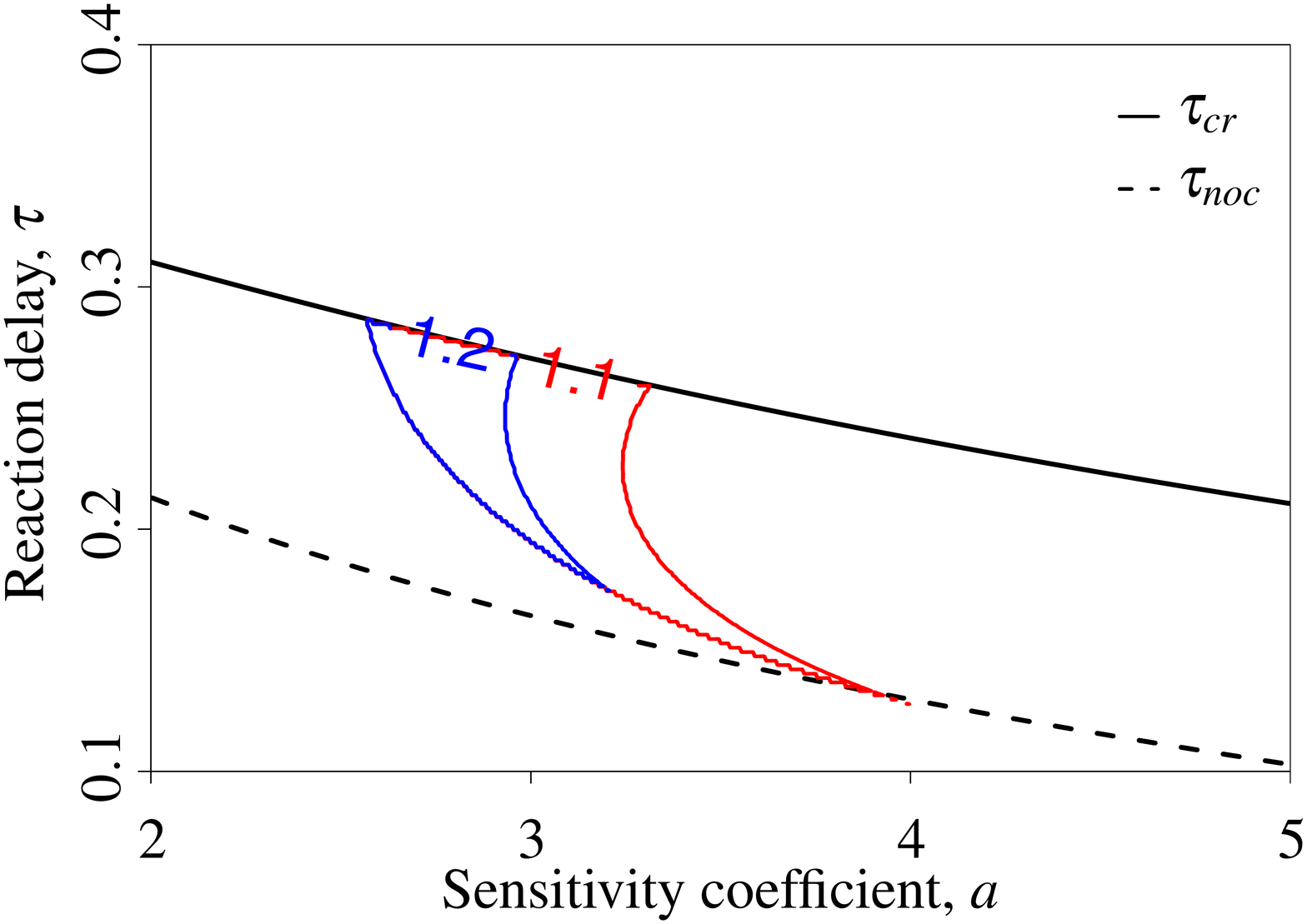}
\label{fig:roccnt2}
  }
\caption{\emph{Contour plots:} Contour lines of the rate of convergence overlaying the boundaries of the locally stable region and the region of non-oscillatory convergence of the MOVM. While $(a)$ is for low to medium values of rate of convergence, $(b)$ is for high values. From $(a),$ observe: $(i)$ the rapid change in the gradient of the rate of convergence, and $(ii)$ for lower values of the rate of convergence, non-oscillatory convergence is also guaranteed. In contrast, $(b)$ shows that very high rates of convergence cannot be achieved if the solutions are to be non oscillatory.} \label{fig:roccnt}
\end{center}
\end{figure*}

The characteristic equation corresponding to the above system is obtained by substituting $\psi = \tau \lambda$ as
\begin{align}
\label{eq:roc_char_eqn}
\lambda^2 + \left( -\frac{2 \sigma}{\tau} \right) \lambda + (a e^{\sigma}) \lambda e^{-\lambda \tau} + \left( d - \frac{a \sigma}{\tau} \right)e^{\sigma} e^{-\lambda \tau} + \left( \frac{\sigma^2}{\tau^2} \right) = 0.
\end{align} 
%\begin{align}
%\label{eq:roc_char_eqn}
%\lambda^2  -\frac{2 \sigma \lambda}{\tau} + a \lambda e^{\sigma -\lambda \tau} + \left( d - \frac{a \sigma}{\tau} \right)e^{\sigma -\lambda \tau} + \frac{\sigma^2}{\tau^2} = 0.
%\end{align}
%The rate of convergence is the largest $\sigma$ such that the root of~\eqref{eq:roc_char_eqn} with the largest real part is negative~\cite{SC}. As pointed out in~\cite{SC}, finding such a $\sigma$ analytically is intractable. Hence, we illustrate this using a numerical example conducted using the scientific computation software MATLAB.

The rate of convergence is the largest $\sigma \geq 0$ such that the root of~\eqref{eq:roc_char_eqn} with the largest real part is negative~\cite{SC}. As pointed out in~\cite{SC}, finding such a $\sigma$ analytically is intractable. Hence, we illustrate the variation of the rate of convergence numerically with respect to both the sensitivity parameter $a$ and the reaction delay $\tau$, using the scientific computation software MATLAB.

We consider the Bando OVF, and set the following parameters: $y_m = 1$ m, $\tilde{y} = 5$ m, $y^{*} = 2$ m and $\dot{x}_0 = 5$ m/s. We then compute the corresponding values of $V_0$ and $\tilde{d}.$ Next, we vary the sensitivity coefficient $a$ in the range $[1,5],$ and for each of its values, we compute the critical value of the reaction delay $\tau_{cr}$ using~\eqref{eq:nscMOVM}. We then vary the reaction delay $\tau$ in the range $[0,\tau_{cr}],$ for each $a.$ For every pair $(a,\tau)$ in this range, $\sigma$ is increased from $0,$ till the root of~\eqref{eq:roc_char_eqn} with the largest real part crosses the imaginary axis in the Argand plane. Since the resulting plot would be three dimensional, we present the corresponding contour plots in Fig.~\ref{fig:roccnt}. For clarity in presentation, the contour plots are segregated as follows: Fig.~\ref{fig:roccnt1} is for low to medium values of the rate of convergence, whereas Fig.~\ref{fig:roccnt2} is for high values. It can be seen from Fig.~\ref{fig:roccnt1} that small changes in $a$ or $\tau$ causes the rate of convergence to change from $0.3$ to $0.9.$ However, it would require relatively larger changes in $a$ or $\tau$ for the rate of convergence to change from $0.1$ to $0.3.$ That is, the gradient of the rate of convergence increases rather rapidly with an increase in the rate of convergence. Also, for low values of the rate of convergence, non-oscillatory convergence can be guaranteed. In contrast, Fig.~\ref{fig:roccnt2} brings forth the trade-off between the rate of convergence and non-oscillatory convergence; very high rates of convergence cannot be achieved if the solutions are to be non oscillatory.

The rate of convergence determines the time taken by a platoon to reach an equilibrium (denoted by $T_{MOVM}^e$). To characterize $T_{MOVM}^e,$ we first define the time taken by the $i^{th}$ pair of vehicles in the platoon following the standard control-theoretic notion of ``settling time.'' That is, by $t_{i}^e(\epsilon),$ we denote the minimum time taken by the time-domain trajectory of the MOVM to enter and subsequently remain within the $\epsilon$-band around the equilibrium. For simplicity, we drop the explicit dependence on $\epsilon.$ Then,
\begin{align}
\label{eq:MOVMequi}
T_{MOVM}^e = \sum_{i = 1}^N t_{i}^e.
\end{align}
It is clear that~\eqref{eq:MOVMequi} is an upper bound on the time taken by the platoon to equilibrate. However, the equality holds since the $i^{th}$ pair cannot equilibrate till the $(i-1)^{th}$ pair has reached its equilibrium.

%% file: sec8_hba.tex
In the previous sections, we have characterized the stable region for the MOVM, and studied two of its most important properties; namely, non-oscillatory convergence and the rate of convergence. We have also proved that system~\eqref{eq:MOVM123} loses stability via a Hopf bifurcation, thus resulting in limit cycles. In this section, we provide an analytical framework to characterize the \emph{type} of the bifurcation and the \emph{asymptotic orbital stability} of the emergent limit cycles. We closely follow the style of analysis presented in~\cite{BDH}, which uses Poincar\'{e} normal forms and the center manifold theory.

We begin by denoting the RHS of~\eqref{eq:modMOVM123eqn} as $f_i.$ That is, for $i \in \lbrace 1, 2, \cdots, N \rbrace,$
\begin{align}
\label{eq:modelrhs}
f_i \triangleq a \kappa \left( V(y_{i-1}(t - \tau_{i-1})) - V(y_{i}(t - \tau_i)) - v_k(t - \tau_i) \right).
\end{align}
%$i \in \lbrace 1, 2, \cdots, N \rbrace.$

Define $\mu = \kappa - \kappa_{cr}.$ Notice that the system undergoes a Hopf bifurcation at $\mu=0,$ where $\kappa = \kappa_{cr}.$ Henceforth, we shall consider $\mu$ as the bifurcation parameter. Also, it is clear that when $\mu >0,$ the exogenous parameter $\kappa$ changes from $\kappa_{cr}$ to $\kappa_{cr} + \mu,$ thus pushing the system into its unstable regime.

We now provide a step-by-step overview of the detailed local bifurcation analysis, before delving into its technical details.

\emph{Step 1}: We begin by applying Taylor's series expansion to the RHS of~\eqref{eq:modelrhs}. Next, we separate the linear terms from their non-linear counterparts. This allows us to cast the resulting equation into the standard form of an Operator Differential Equation (OpDE).

\emph{Step 2}: When $\mu=0,$ the system has exactly one pair of purely imaginary eigenvalues with non-zero angular velocity, as seen from~\eqref{eq:MOVMtranscond}. We call the linear space spanned by the corresponding eigenvectors as the critical eigenspace. For the purpose of our analysis, we also require a locally invariant manifold that is a tangent to the critical eigenspace at the system's equilibrium. The center manifold theorem~\cite{BDH} guarantees the existence of such a manifold.

\emph{Step 3}: Next, we project the system onto its critical eigenspace and its complement when $\mu = 0.$ Thus, we may write the dynamics of the original system on the center manifold as an ODE in a single complex variable.

\emph{Step 4}: Finally, using Poincar\'{e} normal forms, we evaluate the Lyapunov coefficient and the Floquet exponent. These, in turn, help characterize the type of the Hopf bifurcation and the asymptotic orbital stability of the emergent limit cycles.

We begin the analysis by expanding~\eqref{eq:modelrhs} about the equilibrium $v_i^* = 0,$ $y_i^* = V^{'}(V^{-1}(\dot{x}_0)),$ $i = 1, 2, \cdots, N,$ using Taylor's series, to obtain
\begin{align}
\nonumber
\dot{v}_i(t) = & (-\kappa a) v_{i,t}(-\tau_i) + \Omega_i^{(1)} y_{i,t}(-\tau_i)  + \Omega_i^{(2)} y_{i,t}^2(-\tau_i) + \Omega_i^{(3)} y_{i,t}^3(-\tau_i) + \zeta_i^{(1)} y_{(i-1),t}(-\tau_{i-1}) \\ \nonumber
& +  \zeta_i^{(2)} y_{(i-1),t}^2(-\tau_{i-1}) + \zeta_i^{(3)} y_{(i-1),t}^3(-\tau_{i-1}) + \text{higher order terms}, \\
\dot{y}_i(t) = & \, \kappa x_i(t) \label{eq:taylorexpanded}
\end{align}
where we use the shorthand $v_{i,t}(-\tau_i)$ and $y_{i,t}(-\tau_i)$ to represent $v_i(t - \tau_i)$ and $y_i(t - \tau_i)$ respectively. The coefficients are given by $\Omega_i^{(1)} = -\kappa a V^{'}(y_i^*),$ $\Omega_i^{(2)} = -\kappa a V^{''}(y_i^*),$ and $\Omega_i^{(3)} = -\kappa a V^{'''}(y_i^*).$ Also, the coefficients $\zeta_i^{(1)},$ $\zeta_i^{(2)}$ and $\zeta_i^{(3)}$ assume the values $\Omega_i(1),$ $\Omega_i(2)$ and $\Omega_i(3)$ respectively for $i > 1,$ and are zero for $i = 1.$ The terms with the primes $V^{'},$ $V^{''}$ and $V^{'''}$ denote the first, second and third derivatives of the OVF with respect to the state variable respectively.

In the following, we use $\mathcal{C}^k\left(A;B\right)$ to denote the linear space of all functions from $A$ to $B$ which are $k$ times differentiable, with each derivative being continuous. Also, we use $\mathcal{C}$ to denote $\mathcal{C}^0,$ for convenience. 

With the concatenated state $\textbf{S}(t),$ note that~\eqref{eq:modMOVM123eqn} is of the form:
\begin{align}
\label{eq:tryandreduce}
\frac{\text{d}\textbf{S}(t)}{\text{d}t} = \mathcal{L}_{\mu} \textbf{S}_t(\theta) + \mathcal{F}(\textbf{S}_t(\theta), \mu),
\end{align}
where $t > 0$, $\mu \in \mathbb{R}$, and where for $\tau = \underset{i}{\text{max }} \tau_i > 0$,
\begin{align*}
\textbf{S}_t(\theta) = \textbf{S}(t + \theta), \text{ } \textbf{S}: [-\tau,0] \longrightarrow \mathbb{R}^{2N}, \text{ } \theta \in [-\tau, 0].
\end{align*}
Here, $\mathcal{L}_{\mu}: \mathcal{C}\left([-\tau,0];\mathbb{R}^{2N}\right) \longrightarrow \mathbb{R}^{2N}$ is a one-parameter family of continuous, bounded linear functionals, whereas the operator $\mathcal{F}: \mathcal{C}\left([-\tau,0];\mathbb{R}^{2N}\right) \longrightarrow \mathbb{R}^{2N}$ is an aggregation of the non-linear terms. Further, we assume that $\mathcal{F}(\textbf{S}_t, \mu)$ is analytic, and that $\mathcal{F}$ and $\mathcal{L}_{\mu}$ depend analytically on the bifurcation parameter $\mu$, for small $|\mu|$. The objective now is to cast~\eqref{eq:tryandreduce} in the standard form of an OpDE:
\begin{align}
\label{eq:reducetothis}
\frac{\text{d}\textbf{S}_t}{\text{d}t} = \mathcal{A}(\mu) \textbf{S}_t + \mathcal{R} \textbf{S}_t,
\end{align}
since the dependence here is on $\textbf{S}_t$ alone rather than both $\textbf{S}_t$ and $\textbf{S}(t)$. To that end, we begin by transforming the linear problem $\text{d}\textbf{S}(t) / \text{d}t = \mathcal{L}_{\mu} \textbf{S}_t(\theta)$. We note that, by the \emph{Riesz representation theorem}~\cite[Theorem 6.19]{WR}, there exists a $2N \times 2N$ matrix-valued measure $\eta(\cdot, \mu): \mathcal{B}\left(\mathcal{C}\left([-\tau,0];\mathbb{R}^{2N}\right)\right) \longrightarrow \mathbb{R}^{2N \times 2N}$, wherein each component of $\eta(\cdot)$ has bounded variation, and for all $\phi \in \mathcal{C}\left([-\tau,0];\mathbb{R}^{2N}\right),$ we have
\begin{align}
\label{eq:particularleqn}
\mathcal{L}_{\mu} \phi = \int\limits_{-\tau}^0 \text{d}\eta(\theta, \mu) \phi(\theta).
\end{align}
In particular,
\begin{align*}
\mathcal{L}_{\mu} \textbf{S}_t =  \int\limits_{-\tau}^0 \text{d}\eta(\theta, \mu) \textbf{S}(t +\theta).
\end{align*}

Motivated by the linearized system~\eqref{eq:modLMMOVMeqn}, we define
\begin{align*}
\text{d}\eta = 
\begin{bmatrix}
\tilde{A} & \tilde{B} \\
\tilde{C} & \tilde{D}
\end{bmatrix} \text{d} \theta ,
\end{align*}
where \begin{equation*}
(\tilde{A})_{ij} = \begin{cases}
-\kappa d \delta(\theta + \tau_i), & i = j,\\
\kappa d \delta(\theta + \tau_j), & j = i - 1, i > 1,\\
0, & \text{otherwise,}
\end{cases}
\end{equation*}

\begin{equation*}
(\tilde{B})_{ij} = \begin{cases}
-\kappa a \delta(\theta + \tau_i), & i = j,\\
0, & \text{otherwise,}
\end{cases}
\end{equation*}

$\tilde{C} = \kappa I_{N \times N}$ and $\tilde{D} = 0_{N \times N}.$ For instance, when $N$ = 2,
\begin{equation*}
\text{d} \eta = 
\begin{bmatrix}
-\kappa d \delta(\theta + \tau_1) & 0 & -\kappa a \delta(\theta + \tau_1) & 0 \\
\kappa d \delta(\theta + \tau_1) & - \kappa d \delta(\theta + \tau_2) & 0 & -\kappa a \delta(\theta + \tau_2) \\
\kappa & 0 & 0 & 0 \\
0 & \kappa & 0 & 0
\end{bmatrix} \text{d} \theta.
\end{equation*}

For $\phi \in \mathcal{C}^1 \left([-\tau,0];\mathbb{C}^{2N}\right)$, we define
\begin{align}
\label{eq:aoperatordefinition}
\mathcal{A}(\mu) \phi(\theta) = \begin{cases}
\frac{\text{d} \phi(\theta)}{\text{d} \theta}, & \theta \in [-\tau,0),\\
\int\limits_{-\tau}^0 \text{d}\eta(s, \mu) \phi(s) \equiv \mathcal{L}_{\mu}, & \theta = 0,
\end{cases}
\end{align}
and
\begin{align*}
\mathcal{R} \phi(\theta) = \begin{cases}
0, & \theta \in [-\tau,0),\\
\mathcal{F}(\phi, \mu), & \theta = 0.
\end{cases}
\end{align*}

With the above definitions, we observe that $\text{d} \textbf{S}_t / \text{d} \theta \equiv \text{d} \textbf{S}_t / \text{d} t.$ Hence, we have successfully cast~\eqref{eq:tryandreduce} in the form of~\eqref{eq:reducetothis}. To obtain the required coefficients, it is sufficient to evaluate various expressions for $\mu = 0,$ which we use henceforth. We start by finding the eigenvector of the operator $\mathcal{A}(0)$ with eigenvalue $\lambda(0) = j \omega_0$. That is, we want an $2N \times 1$ vector (to be denoted by $q(\theta)$) with the property that $\mathcal{A}(0) q(\theta) = j \omega_0 q(\theta).$ We assume the form: $q(\theta) = [1 \text{ } \phi_1 \text{ } \phi_2 \cdots \text{ } \phi_{2N-1}]^T \text{ } e^{j \omega_0 \theta},$ and solve the eigenvalue equations. That is, we need to solve
\begin{equation*}
\begin{bmatrix}
-\kappa d e^{-j \omega_0 \tau_1} - a \phi_N e^{-j \omega_0 \tau_1} \\
\kappa d e^{-j \omega_0 \tau_2} + \kappa ( -d \phi_1 - a \phi_{N+1} ) e^{-j \omega_0 \tau_2} \\
\kappa d e^{-j \omega_0 \tau_3} + \kappa ( -d \phi_2 - a \phi_{N+2} ) e^{-j \omega_0 \tau_3} \\
\vdots \\
\kappa d e^{-j \omega_0 \tau_{N-1}} + \kappa ( -d \phi_{N-1} - a \phi_{2N-1} ) e^{-j \omega_0 \tau_{N-1}} \\
\kappa \beta \\
\kappa \phi_1 \beta \\
\vdots \\
\kappa \phi_{N-1} \beta
\end{bmatrix} = j \omega_0
\begin{bmatrix}
1 \\
\phi_1 \\
\phi_2 \\
\vdots \\
\phi_{2N-2} \\
\phi_{2N-1}
\end{bmatrix},
\end{equation*}
where $\beta = j (-1 + j e^{-j \omega_0 \tau}) / \omega_0.$ This, in turn, necessitates the following assumptions:
\begin{enumerate}
\item[(i)] \begin{align*}
-\frac{j \omega_0 e^{j \omega_0 \tau_1} + \kappa d}{\kappa^2 a} = \frac{-1 + e^{j \omega_0 \tau}}{\omega_0^2},
\end{align*}
\item[(ii)] For each $i \in \lbrace 1, 2, \cdots N-1 \rbrace,$ the following matrix is invertible:
\begin{align*}
\begin{bmatrix}
\kappa d e^{- j \omega_0 \tau_{i+1}} + j \omega_0 & \kappa a e^{- j \omega_0 \tau_{i+1}} \\
\kappa \beta & - j \omega_0
\end{bmatrix},
\end{align*}
\end{enumerate}
where $\beta = j (-1 + j e^{-j \omega_0 \tau}) / \omega_0.$ Then, for $i \in \lbrace 1, 2, \cdots N-1 \rbrace,$
\begin{align*}
\phi_N = \frac{\kappa \beta}{j \omega_0}, \text{ } \phi_i = \, - \frac{j \kappa \omega_0 d e^{-j \omega_0 \tau_{i}}}{\Delta M_i} , \text{ and } \phi_{N+i}  = \, - \frac{\beta \kappa^2 d e^{-j \omega_0 \tau_{i}}}{\Delta M_i} ,
\end{align*}
where $\Delta M_i = \omega_0^2 - \kappa d \omega_0 \sin (\omega_0 \tau_{i+1}) - \kappa^2 \beta a \cos (\omega_0 \tau_{i+1}) + j \big( \kappa^2 \beta a \sin (\omega_0 \tau_{i+1}) - \kappa d \omega_0 \cos (\omega_0 \tau_{i+1}) \big).$

We define the \emph{adjoint} operator as follows:
\begin{align*}
\mathcal{A}^*(0) \phi(\theta) = \begin{cases}
-\frac{\text{d} \phi(\theta)}{\text{d} \theta}, & \theta \in (0, \tau],\\
\int\limits_{-\tau}^0 \text{d}\eta^T(s, 0) \phi(-s), & \theta = 0,
\end{cases}
\end{align*}
where d$\eta^T$ is the transpose of d$\eta.$
% and is given by
%\begin{equation*}
%(\text{d}\eta^T)_{ij} = \begin{cases}
%-\kappa \beta^*_i \delta(\theta + \tau_i), & i = j,\\
%\kappa \beta^*_{i} \delta(\theta + \tau_{i}), & j = i - 1, j > 1,\\
%0, & \text{otherwise}.
%\end{cases}
%\end{equation*}
We note that the domains of $\mathcal{A}$ and $\mathcal{A}^*$ are $\mathcal{C}^1\left([-\tau,0];\mathbb{C}^{2N}\right)$ and $\mathcal{C}^1\left([0,\tau];\mathbb{C}^{2N}\right)$ respectively. Therefore, if $j \omega_0$ is an eigenvalue of $\mathcal{A},$ then $-j \omega_0$ is an eigenvalue of $\mathcal{A}^*.$ Hence, to find the eigenvector of $\mathcal{A}^*(0)$ corresponding to $-j \omega_0,$ we assume the form: $p(\theta) = B [\psi_{2N-1} \text{ } \psi_{2N-2} \text{ } \psi_{2N-3} \text{ } \cdots \text{ } 1 ]^T \text{ } e^{j \omega_0 \theta},$ and solve $\mathcal{A}^*(0) p(\theta) = - j \omega_0 p(\theta).$ Simplifying this, we obtain
\begin{equation*}
\begin{bmatrix}
-\kappa d e^{-j \omega_0 \tau_1} \psi_{2N-1} + \kappa d e^{-j \omega_0 \tau_1} \psi_{2N-2}  - \beta \kappa \psi_{N-1} \\
-\kappa d e^{-j \omega_0 \tau_2} \psi_{2N-2} + \kappa d e^{-j \omega_0 \tau_2} \psi_{2N-3} - \beta \kappa \psi_{N-2} \\
-\kappa d e^{-j \omega_0 \tau_3} \psi_{2N-3} + \kappa d e^{-j \omega_0 \tau_3} \psi_{2N-4} - \beta \kappa \psi_{N-3} \\
\vdots \\
-\kappa d e^{-j \omega_0 \tau_N} \psi_{N} - \beta \kappa \\
-\kappa a e^{-j \omega_0 \tau_1}  \\
-\kappa a e^{-j \omega_0 \tau_2} \\
\vdots \\
-\kappa a e^{-j \omega_0 \tau_N}
\end{bmatrix} = - j \omega_0
\begin{bmatrix}
\psi_{2N-1} \\
\psi_{2N-2} \\
\psi_{2n-3} \\
\vdots \\
\psi_1 \\
1
\end{bmatrix}.
\end{equation*}
We require the following assumptions:
\begin{enumerate}
\item[(i)] \begin{align*}
\frac{\kappa d - j \omega_0 e^{j \omega_0 \tau_N}}{\kappa^2 a} = \frac{-1 + e^{j \omega_0 \tau}}{\omega_0^2},
\end{align*}
\item[(ii)] For each $i \in \lbrace 1, 2, \cdots N-1 \rbrace,$ the following matrix is invertible:
\begin{align*}
\begin{bmatrix}
j \omega_0 & - \kappa a e^{- j \omega_0 \tau_{i}} \\
\kappa \beta & \kappa d e^{- j \omega_0 \tau_i} - j \omega_0
\end{bmatrix}.
\end{align*}
\end{enumerate}
Then, for $i \in \lbrace 1, 2, \cdots N-1 \rbrace,$ we obtain
\begin{align*}
\psi_N = \frac{j \omega_0 e^{j \omega_0 \tau_N}}{\kappa a}, \text{ } \psi_{N+i} = \frac{j \omega_0 \kappa d \psi_{N+i-1} e^{-j \omega_0 \tau_{N-i}}}{\Delta \tilde{M}_i}, \text{ and } \psi_i = \frac{\kappa^2 a d \psi_{N+i-1} e^{-j \omega_0 \tau_{N-i}}}{\Delta \tilde{M}_i},
\end{align*}
where $\Delta \tilde{M}_i = \omega_0^2 + \kappa d \omega_0 \sin (\omega_0 \tau_{N - i}) + \kappa^2 \beta a \cos (\omega_0 \tau_{N-i}) + j \big( \kappa d \omega_0 \cos (\omega_0 \tau_{N-i}) -\kappa^2 \beta a \sin (\omega_0 \tau_{N-i}) \big).$

The normalization condition for Hopf bifurcation requires that $\langle p, q \rangle$ = $1,$ thus yielding an expression for $B.$

For any $q \in \mathcal{C}\left([-\tau,0];\mathbb{C}^{2N}\right)$ and $p \in \mathcal{C}\left([0, \tau];\mathbb{C}^{2N}\right)$, the inner product is defined as
\begin{align}
\label{eq:innerproductdefinition}
\langle p, q \rangle \triangleq \bar{p} \cdot q - \int\limits_{\theta = - \tau}^0 \int\limits_{\zeta = 0}^{\theta} \bar{p}^T(\zeta - \theta) \text{d}\eta q(\zeta) \text{ d}\zeta,
\end{align}
where the overbar represents the complex conjugate and the $``\cdot"$ represents the regular dot product. The value of $B$ such that the inner product between the eigenvectors of $\mathcal{A}$ and $\mathcal{A}^*$ is unity can be shown to be
\begin{align*}
B = \frac{1}{\zeta_1 + \zeta_2 + \zeta_3 + \zeta_4},
\end{align*}
where
\begin{align*}
\zeta_1 & = \left( \frac{2 e^{j \omega_0 \tau} - e^{j 2 \omega_0 \tau} -1 }{2} \right) \sum\limits_{i = 0}^{N-1} \kappa \psi_{N-i-1} \bar{\phi}_i, \text{ } \zeta_2 = \sum\limits_{i = 0}^{N-1} \left(\frac{ e^{j \omega_0 \tau_{i+1}} - e^{j 2 \omega_0 \tau_{i+1}}}{j \omega_0} \right) \kappa \psi_{2N-1-i} ( a \bar{\phi}_i + d \bar{\phi}_{N+i} ), \\
\zeta_3 & = \sum\limits_{i = 0}^{N-2} \left(\frac{ e^{j 2 \omega_0 \tau_{i+1}} - e^{j  \omega_0 \tau_{i+1}}}{j \omega_0} \right) \kappa d \bar{\phi}_i \psi_{2N-2-i}, \text{ and, } \zeta_4 = \sum\limits_{i = 0}^{2N-1} \psi_{2N-1-i} \bar{\phi}_i.
\end{align*}
In the above, we define $\phi_0 = \psi_0 = 0$ for notational brevity.

For $\textbf{S}_t$, a solution of~\eqref{eq:reducetothis} at $\mu$ $=$ $0,$ we define
\begin{align*}
z(t) = \langle p(\theta), \textbf{S}_t \rangle, \text{ and } \textbf{w}(t, \theta) = \textbf{S}_t(\theta) - 2 \text{Real} ( z(t) q(\theta) ).
\end{align*}
Then, on the center manifold $C_0$, we have $\textbf{w}(t, \theta)$ = $\textbf{w}(z(t), \bar{z}(t), \theta)$, where
\begin{align}
\label{eq:onthemanifold}
\textbf{w}(z(t), \bar{z}(t), \theta) = \textbf{w}_{20}(\theta) \frac{z^2}{2} + \textbf{w}_{02}(\theta) \frac{\bar{z}^2}{2} + \textbf{w}_{11}(\theta) z \bar{z} + \cdots .
\end{align}
Effectively, $z$ and $\bar{z}$ are the local coordinates for $C_0$ in $\mathcal{C}$ in the directions of $p$ and $\bar{p}$ respectively. We note that $\textbf{w}$ is real if $\textbf{S}_t$ is real, and we deal only with real solutions. The existence of the center manifold $C_0$ enables the reduction of~\eqref{eq:reducetothis} to an ODE in a single complex variable on $C_0$. At $\mu$ = 0, the said ODE can be described as
\begin{align}
\nonumber
\dot{z}(t) = & \, \left\langle p, \mathcal{A} \textbf{S}_t + \mathcal{R} \textbf{S}_t \right\rangle, \\ \nonumber
= & \, j \omega_0 z(t) + \bar{p}(0). \mathcal{F}\left( \textbf{w}(z, \bar{z}, \theta) + 2 \text{Real}( z(t) q(\theta) ) \right), \\
= & \, j \omega_0 z(t) + \bar{p}(0). \mathcal{F}_0(z, \bar{z}) \label{eq:odeforz}.
\end{align}
This is written in abbreviated form as
\begin{align}
\label{eq:odeforz2}
\dot{z}(t) = j \omega_0 z(t) + g(z, \bar{z}).
\end{align}
The objective now is to expand $g$ in powers of $z$ and $\bar{z}$. However, this requires $\textbf{w}_{ij}(\theta)$'s from~\eqref{eq:onthemanifold}. Once these are evaluated, the ODE~\eqref{eq:odeforz} for $z$ would be explicit (as given by~\eqref{eq:odeforz2}), where $g$ can be expanded in terms of $z$ and $\bar{z}$ as
\begin{align}
 \label{eq:gexpanded}
g(z, \bar{z}) = \, \bar{p}(0). \mathcal{F}_0(z, \bar{z}) = \, g_{20} \frac{z^2}{2} + g_{02} \frac{\bar{z}^2}{2} + g_{11} z \bar{z} + g_{21} \frac{z^2 \bar{z}}{2} + \cdots .
\end{align}
Next, we write $\dot{\textbf{w}} = \dot{\textbf{S}}_t - \dot{z} q - \dot{\bar{z}} \bar{q}.$ Using~\eqref{eq:reducetothis} and~\eqref{eq:odeforz2}, we then obtain the following ODE:
\begin{align*}
\dot{\textbf{w}} = \begin{cases}
\mathcal{A} \textbf{w} - 2 \text{Real} ( \bar{p}(0). \mathcal{F}_0 q(\theta) ), & \theta \in [-\tau, 0),\\
\mathcal{A} \textbf{w} - 2 \text{Real} ( \bar{p}(0). \mathcal{F}_0 q(0) ) + \mathcal{F}_0, & \theta = 0.
\end{cases}
\end{align*}
This can be re-written using \eqref{eq:onthemanifold} as
\begin{align}
\label{eq:rewrittenwode}
\dot{\textbf{w}} = \mathcal{A} \textbf{w} + H(z, \bar{z}, \theta),
\end{align}
where $H$ can be expanded as
\begin{align}
\label{eq:Hexpansion}
H(z, \bar{z}, \theta) = \, H_{20}(\theta) \frac{z^2}{2} + H_{02}(\theta) \frac{\bar{z}^2}{2} + H_{11}(\theta) z \bar{z} + H_{21}(\theta) \frac{z^2 \bar{z}}{2} + \cdots .
\end{align} 
Near the origin, on the manifold $C_0$, we have $\dot{\textbf{w}} = \textbf{w}_z \dot{z} + \textbf{w}_{\bar{z}} \dot{\bar{z}}.$ Using~\eqref{eq:onthemanifold} and~\eqref{eq:odeforz2} to replace $\textbf{w}_z \dot{z}$ (and their conjugates, by their power series expansion) and equating with~\eqref{eq:rewrittenwode}, we obtain the following operator equations:
\begin{align}
\label{eq:opeqn1}
(2 j \omega_0 - \mathcal{A}) \textbf{w}_{20}(\theta) = & \, H_{20}(\theta), \\ \label{eq:opeqn2}
- \mathcal{A} \textbf{w}_{11} = & \, H_{11}(\theta), \\ \label{eq:opeqn3}
-(2 j \omega_0 + \mathcal{A}) \textbf{w}_{02}(\theta) = & \, H_{02}(\theta).
\end{align}
We start by observing that
\begin{align*}
\textbf{S}_t(\theta) = \, \textbf{w}_{20}(\theta) \frac{z^2}{2} + \textbf{w}_{02}(\theta) \frac{\bar{z}^2}{2} + \textbf{w}_{11}(\theta) z \bar{z} \, + z q(\theta) + \bar{z} \bar{q}(\theta) + \cdots .
\end{align*}

From the Hopf bifurcation analysis~\cite{BDH}, we know that the coefficients of $z^2$, $\bar{z}^2$, $z^2 \bar{z}$, and $z \bar{z}$ terms are used to approximate the system dynamics. Hence, we only retain these terms in the expansions. 

To obtain the effect of non-linearities, we substitute the aforementioned terms appropriately in the non-linear terms of~\eqref{eq:taylorexpanded} and separate the terms as required. Therefore, for each $i \in \lbrace 1, 2, \dots, 2N \rbrace$, we have the non-linearity term to be
\begin{align}
\label{eq:nonlinvector}
\mathcal{F}_i  =  \mathcal{F}_{20i} \frac{z^2}{2} + \mathcal{F}_{02i} \frac{\bar{z}^2}{2} + \mathcal{F}_{11i} z \bar{z} + \mathcal{F}_{21i} \frac{z^2 \bar{z}}{2},
\end{align}
where, for $i \in \lbrace 1, 2, \cdots, N \rbrace,$ the coefficients are given by
\begin{align*}
\mathcal{F}_{20i} & = \Omega_i^{(1)} w_{20i}(-\tau_i) + \zeta_{i-1}^{(1)} w_{20(i-1)}(-\tau_{i-1}), \\
\mathcal{F}_{02i} & = \Omega_i^{(1)} w_{02i}(-\tau_i) + \zeta_{i-1}^{(1)} w_{02(i-1)}(-\tau_{i-1}), \\
\mathcal{F}_{11i} & = \Omega_i^{(1)} w_{11i}(-\tau_i) + \zeta_{i-1}^{(1)} w_{11(i-1)}(-\tau_{i-1}), \\
\mathcal{F}_{21i} & = 2\Omega_i^{(2)} \big( w_{20i}(-\tau_i) e^{j \omega_0 \tau_i} + 2 w_{11i}(-\tau_i) e^{-j \omega_0 \tau_i} \big) \\
& + 2\zeta_{i-1}^{(2)} \big( w_{20(i-1)}(-\tau_{i-1}) e^{j \omega_0 \tau_{i-1}} + 2 w_{11(i-1)}(-\tau_{i-1}) e^{-j \omega_0 \tau_{i-1}} \big).
\end{align*}
For $i \in \lbrace N+1, N+2, \cdots, 2N \rbrace,$ each of these coefficients are zero. This is so, since last $N$ states correspond to the headways who evolution equations are linear.

We represent the vector of non-linearities used in~\eqref{eq:odeforz} as $\mathcal{F}_0 = [\mathcal{F}_1 \text{ } \mathcal{F}_2 \text{ }  \text{ } \cdots \text{ } \mathcal{F}_N]^T.$ Next, we compute $g$ using $\mathcal{F}_0$ as
\begin{align}
 \label{eq:gequation}
g(z, \bar{z}) =  \bar{p}(0).\mathcal{F}_0 = \bar{B} \sum\limits_{l = 1}^N \bar{\psi}_{N-l} \mathcal{F}_l.
\end{align}
Substituting~\eqref{eq:nonlinvector} in~\eqref{eq:gequation}, and comparing with~\eqref{eq:gexpanded}, we obtain
\begin{align}
\label{eq:gxeq}
g_{x} = \bar{B} \sum\limits_{l = 1}^N \bar{\psi}_{N-l} \mathcal{F}_{xl},
\end{align}
where $x \in \lbrace 20, 02, 11, 21 \rbrace.$ Using~\eqref{eq:gxeq}, the corresponding coefficients can be computed. However, computing $g_{21}$ requires $\textbf{w}_{20}(\theta)$ and $\textbf{w}_{11}(\theta).$ Hence, we perform the requisite computation next. For $\theta$ $\in$ $[-\tau,0)$, $H$ can be simplified as
\begin{align*}
H(z, \bar{z}, \theta) & = - \text{Real} \left( \bar{p}(0). \mathcal{F}_0 q(\theta) \right), \\
& = - \left( g_{20} \frac{z^2}{2} + g_{02} \frac{\bar{z}^2}{2} + g_{11} z \bar{z} + \cdots \right) q(\theta) \\
& \hspace*{4mm} - \left( \bar{g}_{20} \frac{\bar{z}^2}{2} + \bar{g}_{02} \frac{z^2}{2} + \bar{g}_{11} z \bar{z} + \cdots \right) \bar{q}(\theta),
\end{align*}
which, when compared with~\eqref{eq:Hexpansion}, yields
\begin{align}
\label{eq:H20eqn}
H_{20}(\theta) & = - g_{20} q(\theta) - \bar{g}_{20} \bar{q}(\theta), \\ \label{eq:H11eqn}
H_{11}(\theta) & = - g_{11} q(\theta) - \bar{g}_{11} \bar{q}(\theta).
\end{align}
From~\eqref{eq:aoperatordefinition},~\eqref{eq:opeqn1} and~\eqref{eq:opeqn2}, we obtain the following ODEs:
\begin{align}
\label{eq:finode1}
\dot{\textbf{w}}_{20}(\theta) = & \, 2 j \omega_0 \textbf{w}_{20}(\theta) + g_{20} q(\theta) + \bar{g}_{02} \bar{q}(\theta), \\ \label{eq:finode2}
\dot{\textbf{w}}_{11}(\theta) = & \, g_{11} q(\theta) + \bar{g}_{11} \bar{q}(\theta).
\end{align}
Solving~\eqref{eq:finode1} and~\eqref{eq:finode2}, we obtain
\begin{align}
\label{eq:w20odesoln}
\textbf{w}_{20}(\theta) = & \, - \frac{g_{20}}{j \omega_0} q(0) e^{j \omega_0 \theta} - \frac{\bar{g}_{02}}{3 j \omega_0} \bar{q}(0) e^{-j \omega_0 \theta} + \textbf{e} \text{ } e^{2 j \omega \theta}, \\ \label{eq:w11odesoln}
\textbf{w}_{11}(\theta) = & \, \frac{g_{11}}{j \omega_0} q(0) e^{j \omega_0 \theta} - \frac{\bar{g}_{11}}{ j \omega_0} \bar{q}(0) e^{-j \omega_0 \theta} + \textbf{f},
\end{align}
for some vectors $\textbf{e}$ and $\textbf{f},$ to be determined.

To that end, we begin by defining the following vector: $\tilde{\mathcal{F}}_{20} \triangleq [\mathcal{F}_{201} \text{ } \mathcal{F}_{202} \text{ } \cdots \text{ } \mathcal{F}_{20N}]^T.$
Equating~\eqref{eq:opeqn1} and~\eqref{eq:H20eqn}, and simplifying, yields the operator equation: $2 j \omega_0 \textbf{e} - \mathcal{A} \left( \textbf{e} \text{ } e^{2 j \omega_0 \theta} \right) = \tilde{\mathcal{F}}_{20}.$ To solve this, we assume that the following matrices to be invertible for each $i \in \lbrace 1, 2, \cdots, N \rbrace,$
\begin{align*}
\begin{bmatrix}
2j \omega_o + \kappa(a+d) e^{-j \omega_0 \tau_i} & \kappa(a+d) e^{-j \omega_0 \tau_i} \\
-\kappa \tau & 2j \omega_o
\end{bmatrix}.
\end{align*}
Then, the operator equation can be simplified to
\begin{align*}
\begin{bmatrix}
\big(2 j \omega_0 + \kappa (a+d) e^{-j \omega_0 \tau_1}\big) \textbf{e}_1 + \kappa (a+d) e^{-j \omega_0 \tau_1} \textbf{e}_{N+1} \\
\big(2 j \omega_0 + \kappa (a+d) e^{-j \omega_0 \tau_2}\big) \textbf{e}_2 + \kappa (a+d) e^{-j \omega_0 \tau_2} \textbf{e}_{N+2} \\
\vdots \\
\big(2 j \omega_0 + \kappa (a+d) e^{-j \omega_0 \tau_N}\big) \textbf{e}_N + \kappa (a+d) e^{-j \omega_0 \tau_N} \textbf{e}_{2N} \\
-\kappa \tau \textbf{e}_1 + 2 j \omega_0 \textbf{e}_{N+1} \\
\vdots \\
-\kappa \tau \textbf{e}_N + 2 j \omega_0 \textbf{e}_{2N} \\
\end{bmatrix}
= \tilde{\mathcal{F}}_{20}.
\end{align*}
This, in turn, yields
\begin{align}
\label{eq:evector}
\textbf{e}_i = \frac{2 j \omega_0 \mathcal{F}_{20i}}{\Delta M_i^*}, \text{ and, } \textbf{e}_{N+i} = \frac{\kappa \tau \mathcal{F}_{20i}}{\Delta M_i^*},
\end{align}
for $i \in \lbrace 1, 2, \cdots, N \rbrace.$ Here, $\Delta M_i^*$ $=$ $- 4 \omega_0^2 + 2 \omega_0 \kappa (a+d) \sin(\omega_0 \tau_i) + \tau \kappa^2 (a+d) \cos(\omega_0 \tau_i)$ $+$ $j \big( 2 \omega_0 \kappa (a+d) \cos(\omega_0 \tau_i) - \tau \kappa^2 (a+d) \sin(\omega_0 \tau_i) \big).$

Next, equating~\eqref{eq:opeqn2} and~\eqref{eq:H11eqn}, and simplifying, we obtain the operator equation $\mathcal{A} \textbf{f} = -\tilde{\mathcal{F}}_{11},$ with $\tilde{\mathcal{F}}_{11} \triangleq [\mathcal{F}_{111} \text{ } \mathcal{F}_{112} \text{ } \cdots \text{ } \mathcal{F}_{11N}]^T.$ On simplification, this equation yields
\begin{align*}
\begin{bmatrix}
- \kappa \tau_1 (a+d) (\textbf{f}_1 + \textbf{f}_{N+1}) \\
- \kappa \tau_2 (a+d) (\textbf{f}_2 + \textbf{f}_{N+2}) \\
\vdots \\
- \kappa \tau_1 (a+d) (\textbf{f}_N + \textbf{f}_{2N}) \\
\kappa \tau \textbf{f}_1 \\
\vdots \\
\kappa \tau \textbf{f}_N \\ 
\end{bmatrix}
= - \tilde{\mathcal{F}}_{11}.
\end{align*}
On solving this, we obtain for $i \in \lbrace 1, 2, \cdots, N \rbrace,$
\begin{align}
\label{eq:fvector}
\textbf{f}_i = 0, \text{ and, } \textbf{f}_{N+i} = \frac{\mathcal{F}_{11i}}{\kappa \tau_i (a+d)}.
\end{align}

Substituting for $\textbf{e}$ and $\textbf{f}$ from~\eqref{eq:evector} and~\eqref{eq:fvector} in~\eqref{eq:w20odesoln} and~\eqref{eq:w11odesoln} respectively, we obtain $\textbf{w}_{20}(\theta)$ and $\textbf{w}_{11}(\theta).$ This, in turn, facilitates the computation of $g_{21}.$ We can then compute
\begin{align*}
 c_1(0)  =  \frac{j}{2 \omega_0} \left( g_{20} g_{11} - 2 | g_{11} |^2 - \frac{1}{3} | g_{02} |^2 \right) + & \frac{g_{21}}{2}, \\
 \alpha^{'}(0) = \text{Real}\left( \frac{\text{d} \lambda}{\text{d} \kappa} \right)_{\kappa = \kappa_{cr} },  \text{ } \mu_2 = - \frac{\text{Real}(c_1(0))}{\alpha^{'}(0)}, \text{ and } & \beta_2 = 2 \text{Real}(c_1(0)).
\end{align*}

Here, $c_1(0)$ is known as the Lyapunov coefficient and $\beta_2$ is the Floquet exponent. It is known from~\cite{BDH} that these quantities are useful since
\begin{enumerate}
\item[$(i)$] If $\mu_2 > 0$, then the bifurcation is \emph{supercritical}, whereas if $\mu_2 < 0$, then the bifurcation is \emph{subcritical}.
\item[$(ii)$] If $\beta_2 > 0$, then the limit cycle is \emph{asymptotically orbitally unstable}, whereas if $\beta_2 < 0$, then the limit cycle is \emph{asymptotically orbitally stable}.
\end{enumerate}

We now present numerically-constructed bifurcation diagrams to gain some insight into the effect of various parameters on the amplitude of the limit cycle.

\subsection*{Bifurcation diagrams}

To obtain bifurcation diagrams, we make use of DDE-BIFTOOL~\cite{KE1,KE2}. We first input system~\eqref{eq:modMOVM123eqn} and their first-order derivatives with respect to the state and delayed state variables to DDE-BIFTOOL. We then set $\kappa = 1$ and initialize the model parameters appropriately. We also fix a range of variation for the bifurcation parameter. DDE-BIFTOOL varies the bifurcation parameter accordingly and finds its critical value. We then increase the value of $\kappa$ and record the amplitude of the resulting limit cycle, thus obtaining the bifurcation diagram. We use the SI units throughout. That is, time will be expressed in ``seconds,'' distance in ``meters,'' velocity in ``meters per second'' and the sensitivity coefficient in ``inverse second.'' For our comparison, we consider two optimal velocity functions; namely, the Bando OVF and the Underwood OVF.

\begin{figure*}[t]
 \begin{center}
 \subfloat[]{ 
%\psfrag{a}{\hspace{-2.15cm}\begin{small} \begin{tabular}{c} Amplitude (relative velocity) \\ \\       \end{tabular} \end{small} }
%  \psfrag{k}{\hspace{-0.2cm}Bifurcation parameter, $\kappa$}
 % \psfrag{y = 1}{\scriptsize$y_i^* = 1$}
%    \psfrag{y = 2}{\scriptsize$y_i^* = 2$}
 %     \psfrag{y = 3}{\scriptsize$y_i^* = 3$}
 % \includegraphics[width=2.05in,height=3.15in,angle=270]{bifurcation_movm_tanh.eps}
 \includegraphics[scale=0.26,angle=90]{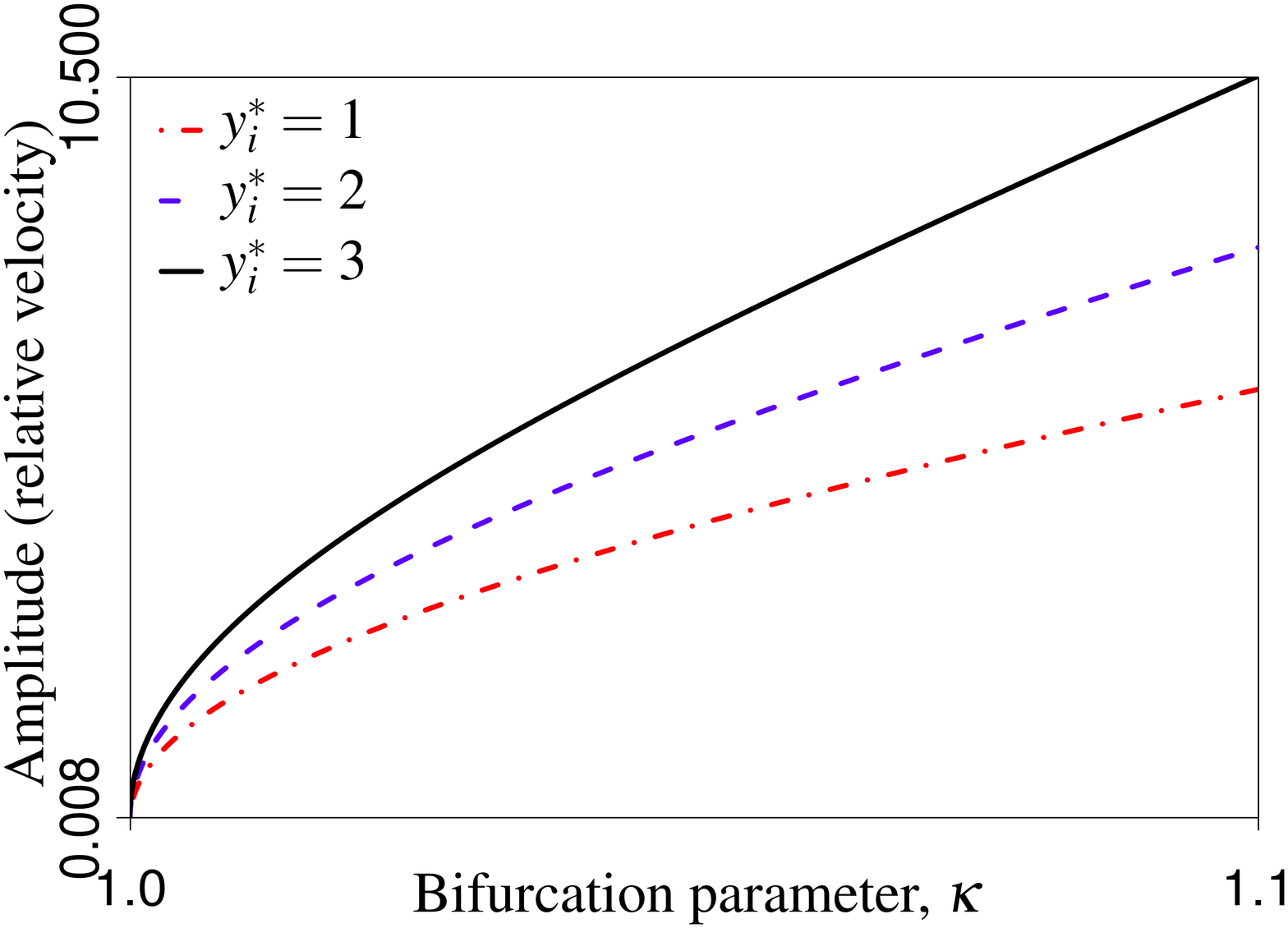}
  \label{fig:bif_diag_tanh}
  } \hspace{1mm}
  \subfloat[]{
%\psfrag{a}{\hspace{-1.1cm}\begin{small} \begin{tabular}{c}Amplitude %(relative velocity) \\ \\       \end{tabular}\end{small} }
%  \psfrag{k}{ \hspace{0.2cm} Bifurcation parameter, $\kappa$}
%  \psfrag{y = 1}{\scriptsize$y_i^* = 1$}
%    \psfrag{y = 2}{\scriptsize$y_i^* = 2$}
%      \psfrag{y = 3}{\scriptsize$y_i^* = 3$}
%  \includegraphics[width=2.05in,height=3.15in,angle=270]{bifurcation_movm_exp.eps}
\includegraphics[scale=0.26,angle=90]{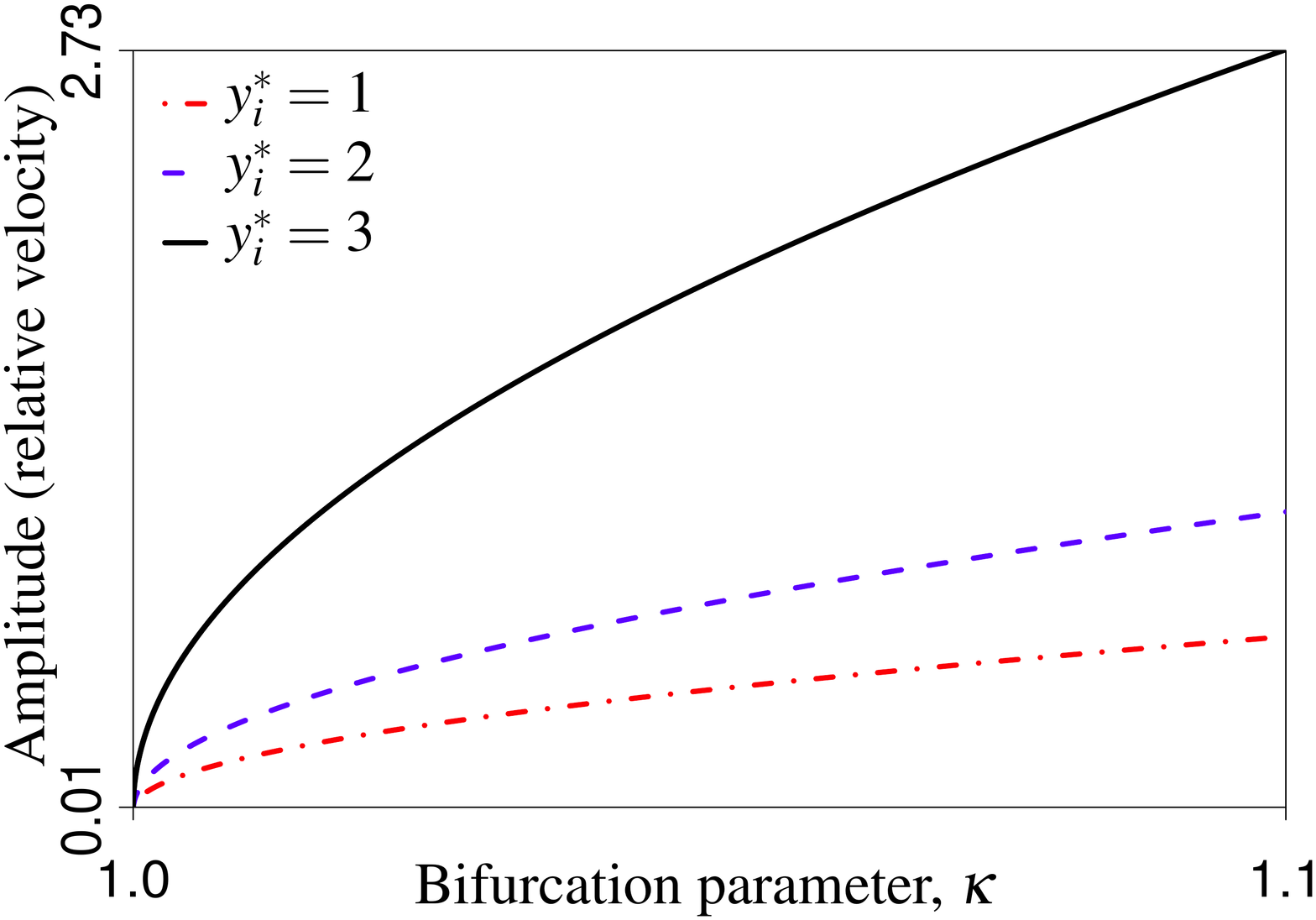}
\label{fig:bif_diag_exp}
  }
\caption{\emph{Bifurcation diagram:} Amplitude of the emergent limit cycles in relative velocity variable as a function of the exogenous parameter $\kappa.$ $(a)$ is for the Bando OVF, while $(b)$ is for the Underwood OVF. For a fixed $\kappa \in [1,1.1],$ the Underwood OVF results in limit cycles of smaller relative velocity than its Bando counterpart.}\label{fig:bif_diag}
\end{center}
\end{figure*}

For the Bando OVF, we initialize the parameters as follows: $N = 4,$ $a = 1.2,$ $\tau_1 = 0.2,$ $\tau_2 = 0.2,$ $\tau_3 = 0.3911$ and $\tau_4 = 0.2.$ We fix $y_m = 2$ and $\tilde{y} = 5,$ and compute $V_0$ for each of $y_i^* = 1, 2$ and $3.$ The vehicle indexed $3$ is considered to undergo a Hopf bifurcation. For the case of the Underwood OVF, we set the following values for the parameters. $N = 3,$ $a = 1.2,$ $\tau_1 = 0.1,$ $\tau_2 = 0.11885$ and $\tau_3 = 0.1.$ We fix $y_m = 2,$ and compute $V_0$ for each of $y_i^* = 1, 2$ and $3.$ The vehicle indexed $2$ is then considered to undergo a Hopf bifurcation. We choose the equilibrium velocity of the lead vehicle, $\dot{x}_0 = 5.$

The bifurcation diagrams are shown in Figure~\ref{fig:bif_diag}. As seen from the figure, the amplitude of the relative velocity increases with an increase in $\kappa.$ However, for a fixed value of the exogenous parameter, the Underwood OVF yields limit cycles with smaller relative velocity that its Bando counterpart, which is desirable. Also, notice that the amplitude of the emergent limit cycles increases with an increase in the equilibrium headway. This is intuitive because larger equilibrium headways offer more space for the resulting limit cycles to oscillate in.

%% file: sec9_sims.tex
Thus far, we have analyzed the MOVM in no-delay, small-delay and arbitrary-delay regimes. We also studied two of its important properties -- non-oscillatory convergence and the rate of convergence. In the previous section, we presented an analytical framework to characterize the type of Hopf bifurcation and the asymptotic orbital stability of the limit cycles that emerge when the stability conditions are marginally violated.

In this section, we present the simulation results of the MOVM that serve to corroborate our analytical findings. We make use of the scientific computation software MATLAB to implement a discrete version of system~\eqref{eq:MOVM123}, thus simulating the MOVM. We use $T_s = 10^{-4}$ s as the update time. Throughout, we use SI units.

\begin{figure*}[t]
 \begin{center}
 \subfloat[]{
 \includegraphics[scale=0.26,angle=90]{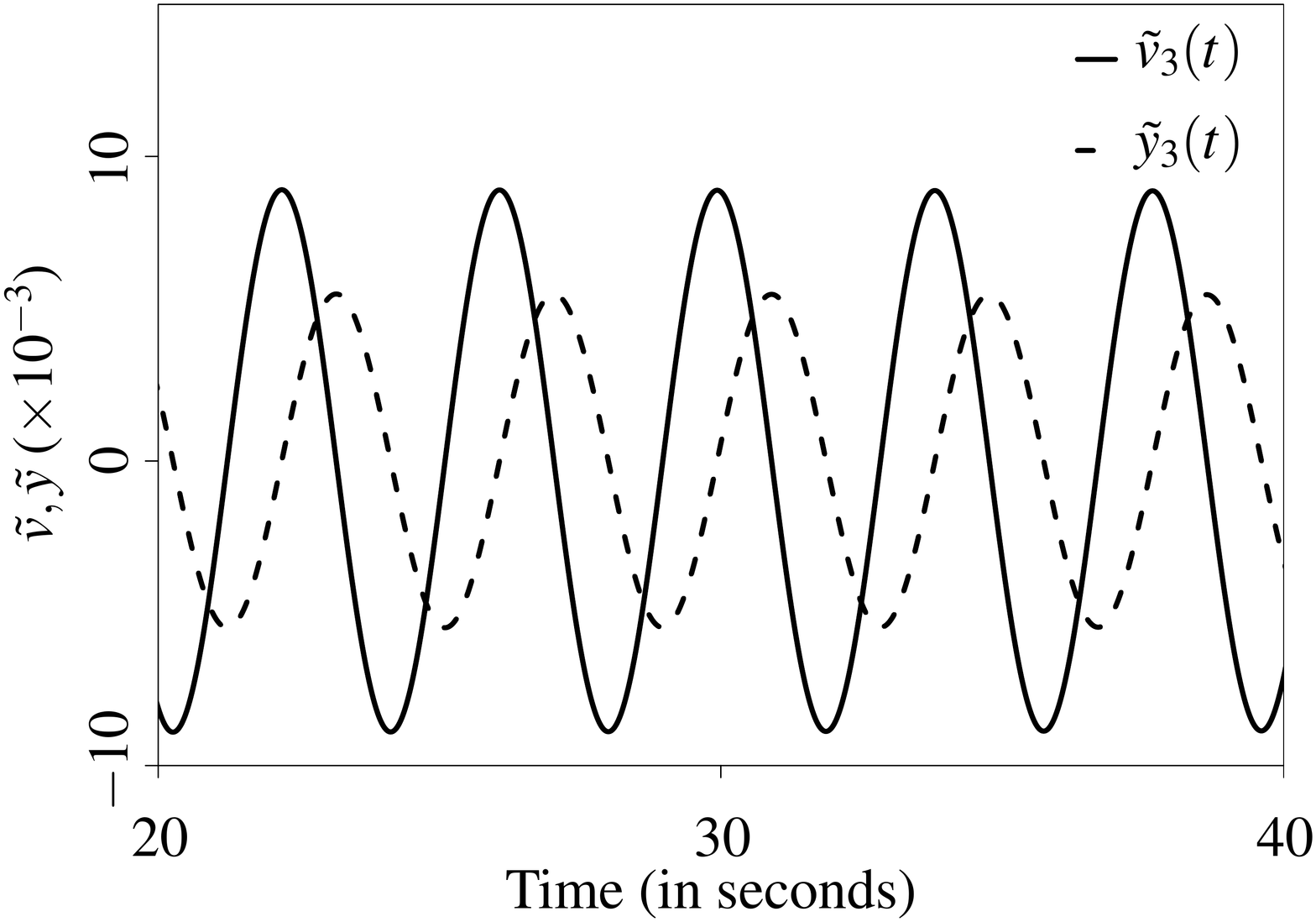}
  \label{fig:sims} 
  } \hspace{1mm}
  \subfloat[]{
\includegraphics[scale=0.26,angle=90]{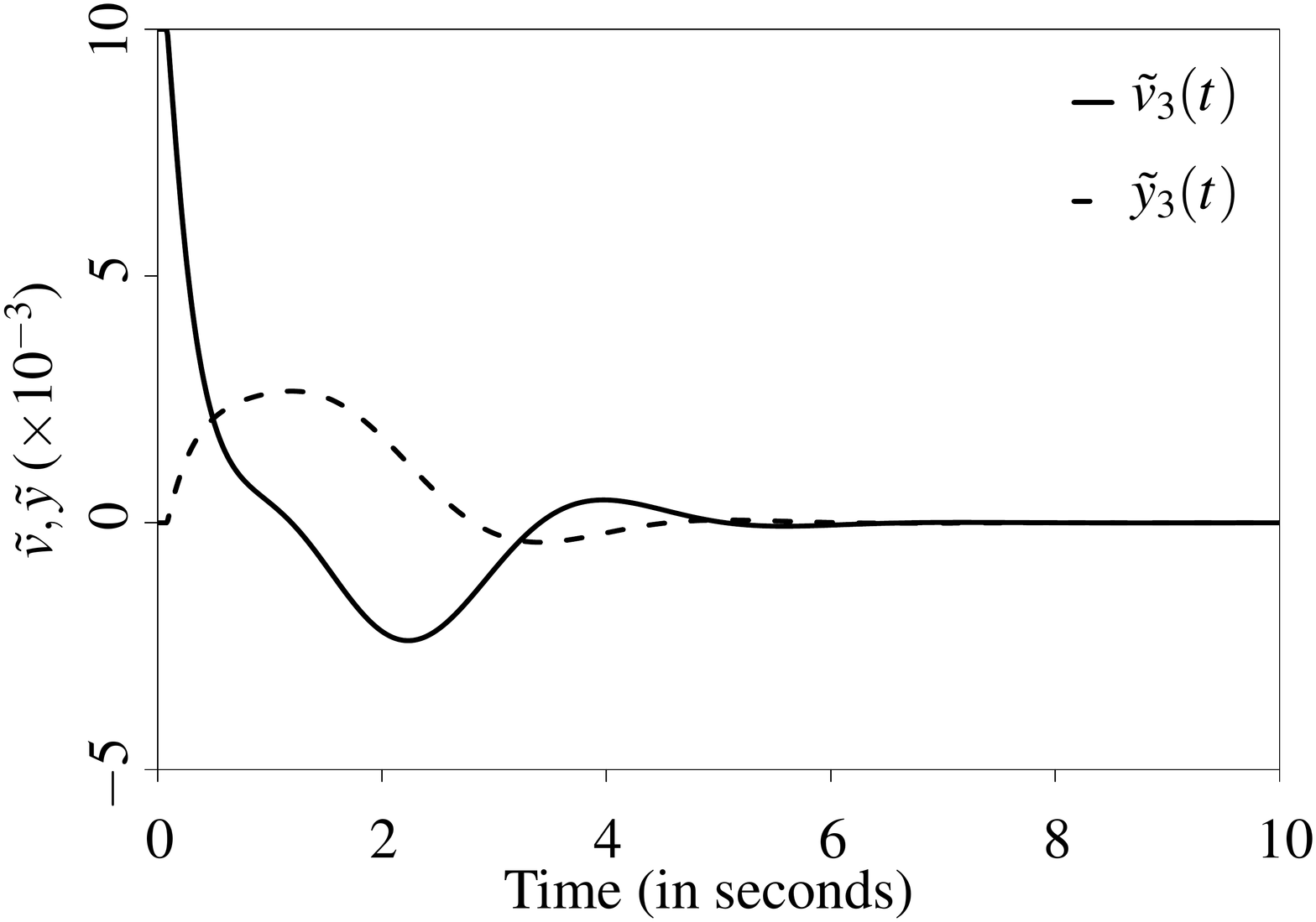}
\label{fig:noc_sims}
  }
\caption{\emph{Simulation results:} Shows the variations in relative velocity and headway around their respective equilibria. $(a)$ portrays the limit cycles predicted by~\eqref{eq:MOVMtranscond}, while $(b)$ presents an instance of non-oscillatory behavior when parameters are chosen appropriately satisfying~\eqref{eq:nscnoc}.}
\end{center}
\end{figure*}

To corroborate the insight from Section~\ref{sec:hopf}, we make use of the Bando OVF. We set the parameters values as: $N = 4,$ $a = 1.2,$ $\tilde{y} = 5,$ $y_m = 1$ and $y_i^{*} = 3$ for $i = 1,2,3,4.$ We then compute the corresponding value of $V_0$ using the functional form for the Bando OVF and $\tau_{cr}$ using~\eqref{eq:nscMOVM}. Further, we set $\tau_1 = \tau_{cr}/10,$ $\tau_2 = \tau_{cr}/3,$ $\tau_3 = \tau_{cr}$ and $\tau_4 = \tau_{cr}/2.$ This ensures that the vehicle indexed $\dot{x}_0 = 3$ undergoes a Hopf bifurcation. The leader's velocity profile is considered to be $5 (1 - e^{-10t}),$ thus yielding an equilibrium velocity of $5.$ We plot the variation of the relative velocity and the headway about their respective equilibria for the vehicle indexed $3.$ That is, we plot $\tilde{v}_3(t) = v_3(t) - v_3^{*} = v_3(t)$ and $\tilde{y}_3(t) = y_3(t) - y_3^{*} = y_3(t) - 3.$ Fig.~\ref{fig:sims} shows the emergence of limit cycles, as predicted by the transversality condition of the Hopf spectrum~\eqref{eq:MOVMtranscond}.

Next, we corroborate the analysis presented in Section~\ref{sec:noc} using the Bando OVF. The parameters are initialized as: $N = 4,$ $a = 2,$ $\tilde{y} = 25,$ $y_m = 15$ and $y_i^{*} = 15$ for $i = 1,2,3,4.$ We then compute the corresponding value of $V_0$ using the functional form for the Bando OVF. We fix the equilibrium velocity of the leader to be $\dot{x}_0 = 25$ by setting the its velocity profile as $25(1 - e^{-10t}).$ We then compute $\tau_{noc}$ using~\eqref{eq:boundaryMOVM}. We set the reaction delays as $\tau_1 = \tau_{noc}/10,$ $\tau_2 = \tau_{noc}/3,$ $\tau_3 = \tau_{noc}/2$ and $\tau_4 = \tau_{noc}/5.$ Fig.~\ref{fig:noc_sims} shows an instance of the relative velocity and headway variations around their respective equilibria for the vehicle indexed $3.$ The headway and relative velocities possess the non-oscillatory behavior, as predicted by the analysis in Section~\ref{sec:noc}.

%% file: sec10_conc.tex
In this paper, we highlighted the importance of delayed feedback in determining the qualitative dynamical properties of a platoon of vehicles traveling on a straight road. Specifically, we analyzed the Modified Optimal Velocity Model (MOVM) in three regimes -- no delay, small delay and arbitrary delay. We proved that, in the absence of reaction delays, the MOVM is locally stable for all practically relevant values of model parameters. We then obtained a sufficient condition for the local stability of the MOVM by analyzing it in the small-delay regime. We also characterized the local stability region of the MOVM in the arbitrary-delay regime.

We then proved that the MOVM undergoes a loss of local stability via a Hopf bifurcation. The resulting limit cycles physically manifest as a back-propagating congestion wave. Thus, our work provides a mathematical basis to explain the observed ``phantom jams.'' For the said analysis, we used an exogenous parameter that captures any interdependence among the model parameters.

We then derived the necessary and sufficient condition for non-oscillatory convergence of the MOVM, with the aim of avoiding jerky vehicular motions. This, in turn, guarantees smooth traffic flow and improves ride quality. Next, we characterized the rate of convergence of the MOVM, which affects the time taken by a platoon to equilibrate. We also brought forth the trade-off between the rate of convergence and non-oscillatory convergence of the MOVM.

Finally, we provided an analytical framework to characterize the type of Hopf bifurcation and the asymptotic orbital stability of the limit cycles which emerge when the stability conditions are violated. Therein, we made use of Poincar\'{e} normal forms and the center manifold theory. We corroborated our analyses using stability charts, bifurcation diagrams, numerical computations and simulations conducted using MATLAB. 

\subsection*{Avenues for further research}

There are numerous avenues that merit further investigation. In this work, we have derived the conditions for pairwise stability of vehicles in a platoon, whose dynamics are captured by the MOVM. However, the string stability of such a platoon remains to be studied.

From a practical standpoint, the parameters of the MOVM may vary, for varied reasons. Hence, it becomes imperative that the longitudinal control algorithm be robust to such parameter variations, and to unmodeled dynamics.

%% file: arxiv_movm.bbl
\begin{thebibliography}{99}

\bibitem{RR}
R. Rajamani, ``Vehicle Dynamics and Control,'' \emph{Springer}, Second Edition, 2012.

\bibitem{AV}
A. Vahidi and A. Eskandarian, ``Research advances in intelligent collision avoidance and adaptive cruise control,'' \emph{IEEE Transactions on Intelligent Transportation Systems}, vol. 4, pp. 143-153 , 2003.

\bibitem{SG}
S. Greengard, ``Smart transportation networks drive gains,'' \emph{Communications of the ACM}, vol. 58, pp. 25-27, 2015.

\bibitem{VAC}
V.A.C. van den Berg and E.T. Verhoef, ``Autonomous cars and dynamic bottleneck congestions: The effects on capacity, value of time and preference heterogeneity,'' \emph{Transportation Research Part B}, vol. 94, pp. 43-60 , 2016.

\bibitem{DCG}
D.C. Gazis, R. Herman and R.W. Rothery, ``Nonlinear follow-the-leader models of traffic flow,'' \emph{Operations Research}, vol. 9, pp. 545-567, 1961.

\bibitem{MBD}
M. Bando, K. Hasebe, K. Nakanishi and A. Nakayama, ``Analysis of optimal velocity model with explicit delay,'' \emph{Physical Review E}, vol. 58, pp. 5429-5435, 1998.

\bibitem{DC}
D. Chowdhury, L. Santen and A. Schadschneider, ``Statistical physics of vehicular traffic and some related systems,'' \emph{Physical Reports}, vol. 329, pp. 199-329, 2000.

\bibitem{DH}
D. Helbing, ``Traffic and related self-driven many-particle systems,'' \emph{Reviews of Modern Physics}, vol. 73, pp. 1067-1141, 2001.

\bibitem{GO}
G. Orosz and G. St$\acute{\text{e}}$p$\acute{\text{a}}$n, ``Subcritical Hopf bifurcations in a car-following model with reaction-time delay,'' \emph{Proceedings of the Royal Society A}, vol. 642, pp. 2643-2670, 2006.

\bibitem{GKK}
G.K. Kamath, K. Jagannathan and G. Raina, ``Car-following models with delayed feedback: local stability and Hopf bifurcation,'' in \emph{Proceedings of the $53^{rd}$ Annual Allerton Conference on Communication, Control and Computing}, 2015.

\bibitem{HL}
J.K. Hale and S.M.V. Lunel, ``Introduction to Functional Differential Equations,'' \emph{Springer-Verlag}, 2011.

\bibitem{RS}
R. Sipahi and S.I. Niculescu, ``Analytical stability study of a deterministic car following model under multiple delay interactions,'' in \emph{Proceedings of Mechanical and Industrial Engineering Faculty Publications}, 2006.

\bibitem{AK}
A. Kesting and M. Treiber, ``How reaction time, update time, and adaptation time influence the stability of traffic flow,'' \emph{Computer-Aided Civil and Infrastructure Engineering}, vol. 23, pp. 125-137, 2008.

\bibitem{MB}
M. Bando, K. Hasebe, A. Nakayama, A. Shibata and Y. Sukiyama, ``Dynamical model of traffic congestion and numerical simulation,'' \emph{Physical Review E}, vol. 51, pp. 1035-1042, 1995.

\bibitem{DCMBD}
L.C. Davis, ``Comment on analysis of optimal velocity model with explicit delay,'' \emph{Physical Review E}, vol. 66, pp. 038101-2, 2002.

\bibitem{LCD}
L.C. Davis, ``Modifications of the optimal velocity traffic model to include delay due to driver reaction time,'' \emph{Physica A}, vol. 319, pp. 557-567, 2003.

\bibitem{IG}
I. Gasser, G. Sirito and B. Werner, ``Bifurcation analysis of a class of `car following' traffic models,'' \emph{Physica D}, vol. 197, pp. 222-241, 2004.

\bibitem{GO1}
G. Orosz, B. Krauskopf and R.E. Wilson, ``Bifurcations and multiple traffic jams in a car-following model with reaction-time delay,'' \emph{Physica D}, vol. 211, pp. 277-293, 2005.

\bibitem{GO2}
G. Orosz, R.E. Wilson and G. St$\acute{\text{e}}$p$\acute{\text{a}}$n, (Editors), ``Traffic jams: dynamics and control,'' \emph{Philosophical Transactions A}, vol. 368, 2010.

\bibitem{REW}
R.E. Wilson and J.A. Ward, ``Car-following models: fifty years of linear stability analysis - a mathematical perspective,'' \emph{Transportation Planning and Technology}, vol. 34, pp. 3-18, 2011.

\bibitem{RR1}
R. Rajamani and C. Zhu, ``Semi-autonomous adaptive cruise control systems,'' \emph{IEEE Transactions on Vehicular Technology}, vol. 51, pp. 1186-1192, 2002.

\bibitem{ZQ}
Z. Qu, J. Wang and R.A. Hull, ``Cooperative control of dynamical systems with application to autonomous vehicles,'' \emph{IEEE Transactions on Automatic Control}, vol. 53, pp. 894-911, 2008.

\bibitem{RUC}
R.U. Chavan, M. Belur, D. Chakraborty and D. Manjunath, ``On the stability and formations in ad hoc multilane vehicular traffic,'' in \emph{Proceedings of the $7^{th}$ International Conference on Communication Systems and Networks (COMSNETS)}, 2015.

\bibitem{THS}
T.H. Summers, C. Yu, S. Dasgupta and B.D.O. Anderson, ``Control of minimally persistent leader-remote-follower and coleader formations in the plane,'' \emph{IEEE Transactions on Automatic Control}, vol. 56, pp. 2778-2792 , 2011.

\bibitem{KCD}
K.C. Dey, L. Yan, X. Wang, Y. Wang, H. Shen, M. Chowdhury, L. Yu, C. Qiu and V. Soundararaj, ``A review of communication, driver characteristics, and controls aspects of cooperative adaptive cruise control (CACC),'' \emph{IEEE Transactions on Intelligent Transportation Systems}, vol. 17, pp. 491-509, 2016.

\bibitem{MTS}
M. Won, T. Park and S.H. Son, ``Toward mitigating phantom jam using vehicle-to-vehicle communication,'' \emph{IEEE Transactions on Intelligent Transportation Systems}, Early Access, 2016.

\bibitem{DC1}
D. Chen, J. Laval, Z. Zheng and S. Ahn, ``A behavioral car-following model that captures traffic oscillations,'' \emph{Transportation Research Part B}, vol. 46, pp. 744-761, 2012.

\bibitem{RN}
R. Nishi, A. Tomoeda, K. Shimura and K. Nishinari, ``Theory of jam-absorbing driving,'' \emph{Transportation Research Part B}, vol. 50, pp. 116-129, 2013.

\bibitem{YT}
Y. Taniguchi, R. Nishi, T. Ezaki and K. Nishinari, ``Jam-absorption driving with a car-following model,'' \emph{Physica A}, vol. 433, pp. 304-315, 2015.

\bibitem{GO3}
G. Orosz, ``Connected cruise control: modelling, delay effects, and nonlinear behaviour,'' \emph{Vehicle System Dynamics}, vol. 54, pp. 1147-1176, 2016.

\bibitem{MAS}
M.~di~Bernardo,~A.~Salvi and~S.~Santini, ``Distributed consensus strategy for platooning of vehicles in the presence of time-varying heterogeneous communication delays,'' \emph{IEEE Transactions on Intelligent Transportation Systems}, vol. 16, pp. 102-112, 2015.

\bibitem{MBET}
M. Batista and E. Twrdy, ``Optimal velocity functions for car-following models,'' \emph{Journal of Zhejiang University - Science A (Applied Physics $\&$ Engineering)}, vol. 11, pp. 520-529, 2010.

\bibitem{GKKITS}
G.K. Kamath, K. Jagannathan and G. Raina, ``A computational study of a variant of the Optimal Velocity Model with no collisions,'' in \emph{Proceedings of $8^{th}$ International Conference on Communication Systems and Networks (COMSNETS)}, 2016.

\bibitem{GL}
I. Gy$\ddot{\text{o}}$ri and G. Ladas, ``Oscillation Theory of Delay Differential Equations With Applications,'' \emph{Clarendon Press}, 1991.

\bibitem{JS}
J.R. Silvester, ``Determinants of block matrices,'' \emph{The Mathematical Gazette}, vol. 84, pp. 460-467, 2000.

\bibitem{BDH}
B.D. Hassard, N.D. Kazarinoff and Y.-H. Wan, ``Theory and Applications of Hopf Bifurcation.,'' \emph{Cambridge University Press}, 1981.

\bibitem{GR}
G. Raina, ``Local bifurcation analysis of some dual congestion control algorithms,'' \emph{IEEE Transactions on Automatic Control}, vol. 50, pp. 1135-1146, 2005.

\bibitem{SM}
S. Manjunath and G. Raina, ``FAST TCP: some queueing models and stability,'' in \emph{Proceedings of International Conference on Signal Processing and Communications (SPCOM)}, 2014.

\bibitem{SC}
S. Chong, S. Lee and S. Kang, ``A simple, scalable, and stable explicit rate allocation algorithm for $\max-\min$ flow control with minimum rate guarantee,'' \emph{IEEE/ACM Transactions on Networking}, vol. 9, pp. 322-335, 2001. 

\bibitem{WR}
W. Rudin, ``Real $\&$ Complex Analysis,'' \emph{Tata McGraw Hill Publications}, Third Edition, 1987.

\bibitem{KE1}
K. Engelborghs, T. Luzyanina and D. Roose, ``Numerical bifurcation analysis of delay differential equations DDE-BIFTOOL,'' \emph{ACM Transactions on Mathematical Software (TOMS)}, vol. 28, pp. 1-21, 2002.

\bibitem{KE2}
K. Engelborghs, T. Luzyanina and G. Samaey, ``DDE-BIFTOOL v. 2.00: a Matlab package for bifurcation analysis of delay differential equations,'' \emph{Technical Report TW-330}, Department of Computer Science, Katholieke Universiteit Leuven, Belgium, 2001.

\end{thebibliography}
